\documentclass[fleqn,usenatbib]{mnras}

\usepackage[T1]{fontenc}

\DeclareRobustCommand{\VAN}[3]{#2}
\let\VANthebibliography\thebibliography
\def\thebibliography{\DeclareRobustCommand{\VAN}[3]{##3}\VANthebibliography}

\usepackage{graphicx}	% Including figure files
\usepackage{amsmath}	% Advanced maths commands
\usepackage{amssymb}	% Extra maths symbols
\usepackage{lscape} 
\usepackage{rotating}
\usepackage{lineno}
\usepackage{mathtools}
\usepackage{newtxtext,newtxmath}
\newcommand*{\logTen}{\ensuremath{\log_{10}}}

\newcommand{\fcont}{{$f_\mathrm{cont}$}}
\title[The RASS-MCMF cluster catalog]{RASS-MCMF: A full-sky X-ray selected galaxy cluster catalog}
\author[Matthias Klein et al.]{
Matthias Klein,$^{1}$\thanks{E-mail:matthias.klein@physik.lmu.de}
Daniel Hern\'andez-Lang,$^{1,2}$
Joseph J Mohr,$^{1,3}$
Sebastian Bocquet,$^{1}$
and Aditya Singh$^{1,3}$
\\
$^{1}$ Ludwig-Maximilians-Universit\"at in Munich, Faculty of Physics, Scheinerstrasse 1, 81679 Munich, Germany\\
$^{2}$ Excellence Cluster Origins, Boltzmannstr.\ 2, 85748 Garching, Germany\\
$^{3}$ Max Planck Institute for Extraterrestrial Physics, Giessenbachstrasse 1, 85748 Garching, Germany\\
}

\date{Accepted XXX. Received YYY; in original form ZZZ}

\pubyear{2023}

\begin{document}
\label{firstpage}
\pagerange{\pageref{firstpage}--\pageref{lastpage}}
\maketitle

\begin{abstract}
We present the RASS-MCMF catalog of 8,465 X-ray selected galaxy clusters over 25,000 deg$^2$ of extragalactic sky.  The accumulation of deep, multiband optical imaging data, the development of the Multi-Component Matched Filter cluster confirmation algorithm (MCMF), and the release of the DESI Legacy Survey DR10 catalog makes it possible-- for the first time, more than 30 years after the launch of the ROSAT X-ray satellite-- to identify the majority of the galaxy clusters detected in the second ROSAT All-Sky-Survey (RASS) source catalog (2RXS).  The resulting 90\% pure RASS-MCMF catalog is the largest ICM-selected cluster sample to date. 
RASS-MCMF probes a large dynamic range in cluster mass spanning from galaxy groups to the most massive clusters.
The cluster redshift distribution peaks at $z\sim0.1$ and extends to redshifts $z\sim1$. Out to $z\sim0.4$, the RASS-MCMF sample contains more clusters per redshift interval ($dN/dz$) than any other ICM-selected sample.
In addition to the main sample, we present two subsamples with 6,924 and 5,516 clusters, exhibiting 95\% and 99\% purity, respectively.   
We forecast the utility of the sample for a cluster cosmological study, using realistic mock catalogs that incorporate most observational effects, including the X-ray exposure time and background variations, the existence likelihood selection and the impact of the optical cleaning with the algorithm MCMF. Using realistic priors on the observable--mass relation parameters from a DES-based weak lensing analysis, we estimate the constraining power of the RASS-MCMF$\times$DES sample to be of 0.026, 0.033 and 0.15 ($1\sigma$) on the parameters $\Omega_\mathrm{m}$, $\sigma_8$ and $w$, respectively.
\end{abstract}



%%%%%%%%%%%%%%%%%%%%%%%%%%%%%%%%%%%%%%%%%%%%%%%%%%

%%%%%%%%%%%%%%%%% BODY OF PAPER %%%%%%%%%%%%%%%%%%

\section{Introduction}
\label{sec:Introduction}

Selecting galaxy clusters through their intracluster medium (ICM) signatures-- either X-ray emission \citep[e.g.,][]{Sarazin88} or the thermal Sunyaev-Zel'dovich effect \citep[SZE; ][]{Sunyaev72}-- is an efficient way to create cluster samples that can be employed for cosmological analyses \citep[e.g.,][]{Vikhlinin98,Mantz10,PlanckSZcosmology}.  X-ray and SZE signatures are dominated by processes in the hot and dense cluster virial regions, which ensures that the distribution of clusters in observable and redshift-- the so-called halo observable function (HOF)-- can be related to the underlying halo mass function (HMF) through observable--mass relations.  An accurate mapping from HOF to HMF is crucial for carrying out cosmological studies using the abundance of galaxy clusters \citep[e.g., ][]{Hu03,Majumdar04,Lima05}.
%,Pratt2019SSRv..215...25P}.

In contrast to ICM selection, selecting galaxy clusters through their passive galaxy populations-- the so-called red sequence methods \citep{gladders00,Rykoff14}-- relies upon a cluster signature that traces not only the dense cluster virial regions but also the low density regions outside cluster and group halos.  Neither photometric redshifts, galaxy colors nor spectroscopic redshifts can be used to identify whether galaxies along the line of sight toward the cluster lie within the cluster virial region or in the surrounding region that extends 10 to 20~Mpc behind and in front of the cluster \citep[e.g.,][]{Song12,Saro13}.  This additional ``contrast'' challenge complicates the interpretation of the number of cluster galaxies-- the richness $\lambda$-- and its relationship to halo mass, and it may also weaken the required one-to-one relationship between optically selected clusters and collapsed halos, making it more difficult to use optically selected cluster abundance to study cosmology.  Methods are being developed to overcome these challenges and have been employed to deliver cosmological constraints \citep[e.g.,][]{costanzi19, CAMIRA-WL, DESY1clucosmo, Lesci2022A&A...659A..88L}.

ICM-selected cluster samples have to be followed up optically to determine the cluster redshifts.  With overlapping deep, multi-band surveys \citep[e.g., KiDS, DES and HSC-SSP;][]{deJong,Flaugher15,HSC-SSP} it is possible to do much more.  One can use the richness of the optical counterpart of an ICM selected cluster to exclude those cluster candidates with low significance optical counterparts, because they are likely contamination \citep{Klein18}.  The ``contrast'' challenge mentioned above has no impact on this process.  Thus, the multi-band survey data allow one to make the most of an X-ray or SZE survey, because one can include ICM-selected counterparts with lower ICM detection significance without increasing the contamination fraction of the final cluster sample.  A benefit of this approach is that lower-mass clusters are included at all redshifts, and the maximum redshift probed by the sample is increased.

Optical followup based on the passive galaxy population has been shown to be robust for clusters and for low redshift high mass groups.  
For systems with $M_{500}\ge3\times10^{14}M_\odot$, purely ICM-selected cluster samples from, e.g., SPT \citep{Carlstrom11}, exhibit dominant passive galaxy populations out to redshifts $z\sim1$ \citep{Hennig17}, and deep Spitzer and HST studies of five of the highest redshift SZE selected clusters from SPT at $1.4<z<1.72$ show higher passive fractions than the field at comparable redshift, indicating environmental quenching efficiencies in the range of 0.5 to 0.8 \citep{Strazzullo19}.  Moreover, in the recent eFEDS X-ray study with eROSITA \citep{Predehl21}, extended sources with masses $M_{500}>5\times10^{13}M_\odot$ exhibit passive dominated galaxy counterparts over the redshifts range that they are sampled \citep{Klein22}.

In addition to redshift estimation, data from deep, multiband surveys like KiDS, DES and HSC-SSP enable one to use the weak gravitational lensing shear and photometric redshift measurements of background galaxies to directly constrain the halo mass distribution of the clusters.  In general, these survey weak lensing datasets are created with the goal of carrying out cosmic shear studies and are therefore more homogeneous with better understood systematics than the pointed cluster weak lensing datasets that have been employed for cluster cosmology in the past.  
%{\bf This allows for more robust cluster mass constraints and will presumably help to avoid situations like the controversial hydrostatic mass bias of the Planck SZE selected cluster sample where multiple, pointed cluster weak-lensing analyses led to some inconsistent mass constraints \citep[see cluster cosmology discussion in][]{ PlanckCollaboration2020overviewlegacy}.}

Larger, high-purity ICM-selected cluster samples extending over a broader mass and redshift range together with weak-lensing mass information on the full sample enable more accurate and precise cosmological studies.  Thus, the combination of X-ray or SZE cluster surveys with deep and homogeneous multi-band optical survey data offers the promise to produce the most constraining cluster cosmological studies to date.

Initial examples of this approach are now emerging.
The optical followup of ACT and SPT-selected cluster candidates already heavily relies on survey data from DES \citep{Bleem2020ApJS..247...25B, ACTDR5}.
New and dramatically larger ICM-selected cluster samples have been produced using the Multi-Component Matched Filter (MCMF) technique.  An analysis of the ROSAT All-Sky X-ray Survey (RASS) together with the DES led to a sample of $\sim$2,000 X-ray selected clusters extending to redshift $z\sim1$ with a surface density that is an order of magnitude higher than that of past RASS cluster catalogs over the same sky region \citep{Klein19}.  An analysis of the Planck SZE selected cluster candidate sample to lower signal to noise \citep{PlanckSZE14} in combination with DES led to a factor of four increase in the number of confirmed clusters over the DES region and extended the maximum redshift of the Planck cluster sample to $z\sim1$ \citep{Hernandez22}.  A similar analysis of the SPT-SZ and SPTpol 500d SZE surveys in combination with DES has also led to a significant increase in the mass range of that SZE selected sample (Klein et al., Bleem et al., both in prep.).  

At present no DES weak-lensing informed cosmological analysis of these enhanced samples has been presented, but several are underway.  However, in the case of the eROSITA eFEDS pilot X-ray survey, the cosmological analysis of the cluster sample has been carried out in combination with the weak-lensing dataset of the HSC-SSP \citep{Chiu23}.  
%A similar analysis using the extended SPT cluster sample mentioned above in combination with DES weak lensing is currently approaching completion (Bocquet et al, in prep).

These efforts provide evidence of the benefits of combining ICM-selected samples with large solid angle, deep multi-band optical surveys.  In this paper we employ the MCMF tool in combination with the latest reanalysis of the all-sky X-ray survey RASS \citep[2RXS;][]{Boller16} and the latest release of the optical and IR multi-band Legacy Survey DR10 \citep[][in prep.]{Legacysurveys19} to produce a new X-ray selected cluster sample called RASS-MCMF.  Our multi-wavelength analysis extends over the bulk of the extragalactic sky, a region of over 25,000~deg$^2$, and yields a high-purity sample of over 8,000 X-ray selected galaxy clusters with redshifts, richnesses, optical centers and X-ray fluxes.

In Section~\ref{sec:Data} we present the data used in this analysis.  Section~\ref{sec:Method} contains a summary of the methods used for cluster catalog construction, and Section~\ref{sec:Catalog} presents the RASS-MCMF cluster sample.  The results of cross-comparison to other cluster samples is presented in Section~\ref{sec:Validation}, while Section~\ref{sec:Forecast} contains a cosmological forecast that highlights the usefulness of the sample.  Conclusions are presented in Section~\ref{sec:Conclusions}.
Throughout this paper we assume a flat $\Lambda$CDM model with $\Omega_\mathrm{m}=0.3$ and $H_0=70$~km$\,$s$^{-1}\,$Mpc$^{-1}$ unless otherwise stated..

\section{Data}
\label{sec:Data}

In the following subsections we describe the X-ray and multi-band optical and IR datasets used for this analysis.

\begin{figure*}
\includegraphics[keepaspectratio=true,width=0.99\linewidth]{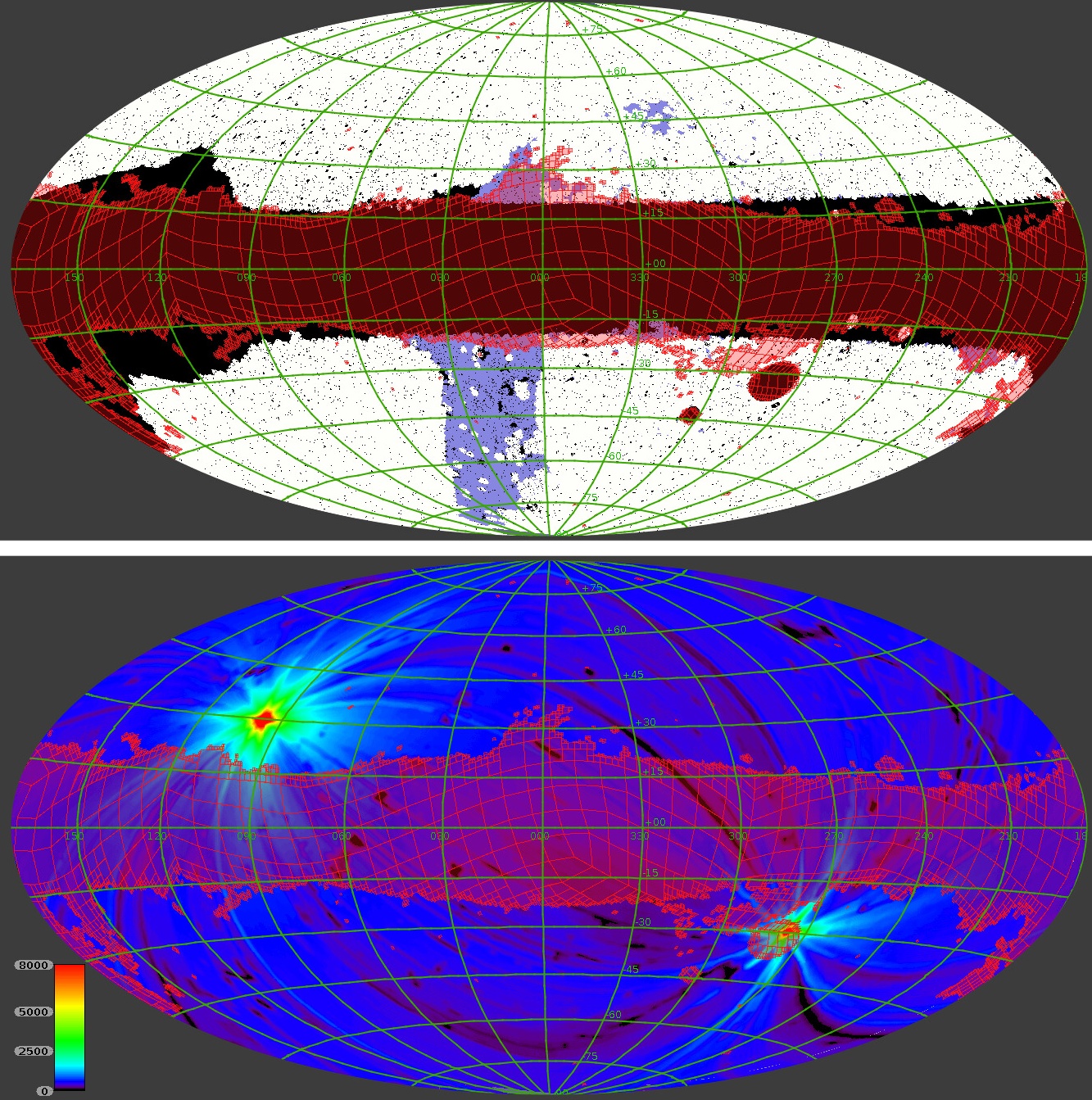}
\vskip-0.10in
\caption{Top: The DESI Legacy Survey DR10 showing extragalactic coverage in $grz$ (white) and $gi$ (blue).  Both datasets are supplemented with WISE $w1$ and $w2$ photometry. Bottom: The ROSAT All-Sky Survey exposure map from the 2RXS source catalog is drawn. Highlighted in red are sky regions impacted by high galactic NH column densities ($N_H>10^{21}$ cm$^{-2}$) or high stellar densities and therefore less suited for cluster search.}
\label{fig:footprint}
\end{figure*}

\subsection{The Second ROSAT All-Sky Survey Source catalog}\label{sec:2rxs}
The ROSAT All-Sky Survey (RASS) was performed more than thirty years ago with the
ROSAT satellite \citep{Truemper93} from June 1990 to August 1991. It was the first all-sky imaging survey in X-rays, resulting in an increase in known X-ray sources by a factor of $\sim$100 \citep{voges99}.
The second ROSAT All-Sky-Survey source catalog \citep[2RXS;][]{Boller16}{}{} builds 
upon previous work \citep[1RXS][]{voges99,voges00}{}{} and uses the more recent RASS-3 processed photon event files together with an improved source detection algorithm resulting in a catalog of 135,000 X-ray sources. Dedicated simulations were performed to estimate the contamination by spurious sources as a function of existence likelihood (EXI\_ML). The released 2RXS catalog is expected to include $\sim$30\% spurious sources and contains all sources with existence likelihood EXI\_ML$\ge$6.5. We expect from previous studies (e.g., Table 2 in \citet{Hasinger96} and also \citet{Klein19}) that $\sim 10-15\%$ of the X-ray sources in extragalactic regions are likely galaxy clusters. This suggest that 2RXS might include approximately $10,000$ groups and clusters. Given the typically low signal to noise of 2RXS sources together with the poor RASS angular resolution of $\sim$4~arcmin \citep{boese00}, a clean and complete selection of these 2RXS detected clusters is not possible from ROSAT data alone.
Figure~\ref{fig:footprint} shows the sky coverage of the RASS data used in this analysis.  The primary 2RXS inputs for the analysis that follows are the source positions and X-ray count rates.

\subsection{DESI Legacy Survey DR10}\label{sec:datals}
The DESI Legacy Imaging Surveys \citep[LS;][]{Legacysurveys19} up to data release 8, was a combination of four imaging surveys, the 9,000 $\text{deg}^2$ $grz$-band DECam \citep{Flaugher15} based DECaLS survey, the 5,000 $\text{deg}^2$ BASS and MzLS surveys providing photometry in $g,r$ and $z$-band, respectively, and the WISE and NEOWISE surveys in the mid-IR at 3.4$\mu$m and 4.6$\mu$m. With the subsequent data releases other DECam based imaging has been included. The most recent data release, DR10, includes imaging data from the Dark Energy Survey as well as from various other survey programs such as BLISS and the DeROSITAS survey. While the BLISS program focuses on imaging the complete DECam-observable sky, the focus of the DeROSITAS survey was to obtain imaging data to enable cluster identification for the portion of the eROSITA X-ray survey that lies within the western galactic hemisphere (the so-called German portion of the eROSITA sky). Therefore a special focus was put on image quality and depth;
experience gained from previous studies of 2RXS sources over DES \citep{Klein19} helped define the DeROSITAS survey parameters and observing plan. 

The recent Legacy DR10 (Dey et al., in prep.) is the addition of DECam imaging data from the $i$-band, mostly coming from the DeROSITAS and DES surveys. The imaging depth depends on sky position, given the differing requirements of the various surveys. The 5,000~deg$^2$ BASS and MzLS surveys, which we call Legacy Survey North, exhibits a typical 5$\sigma$ point source depth of $\sim$24.3, $\sim$23.8 and $\sim$23.4~mag in the $grz$ bands, respectively. The DECam based surveys show typically a double peaked imaging depth distribution \footnote{See https://www.legacysurvey.org/dr10/description/}, the shallower peak is mostly associated with the DECaLS survey with depths of $\sim$24.8, $\sim$24.2 and $\sim$23.3~mag ($grz$) and the deeper peak with the DES survey with depths of $\sim$25.3, $\sim$25.0 and $\sim$23.9~mag. In all cases we consider the imaging depth to be sufficient for the identification of clusters in 2RXS. 

The combination of the optical data with the most recent WISE data allows for improved redshift constraints and increased redshift range due to the strong dependency of the $z-w1$ or $i-w1$ color on redshift. In the top panel of Figure~\ref{fig:footprint} we show the coverage of the DESI Legacy Survey DR10, split into two regions, those containing $grz$ optical imaging (white) and those with $gi$ imaging  with no $z$ band (blue).  The union of both regions is employed for  
galaxy cluster identification in this analysis. While the area covered with $gi$ imaging is as large as $\sim$17,000~deg$^2$ we only consider cluster measurements using the $gi$ dataset within the blue region, which corresponds to $\sim$1,700~deg$^2$.  Outside this region we employ the $grz$ imaging for cluster studies.  The combined footprint is $\sim25,000$~deg$^2$, covering the majority of the extragalactic sky outside a Galactic latitude of $\pm$17~deg. 

Because the Legacy Survey North was conducted with different instruments than that of the rest of the survey, slight differences between colors are expected. 
We calibrate observed galaxy colors and magnitudes from the Legacy Survey North to the Legacy Survey South system \citep[see, e.g.,][]{Duncan22}. Furthermore, we first treat the cluster analysis of the Legacy Survey North region independently, and only merge it with the DECam based dataset after cross checks have been performed.

%%%%%%%%%%%%%%%%%%%%%%%%%%%%%%%%%%%%%%%%%%%%%%%%%%
%%%%% MCMF Method
%%%%%%%%%%%%%%%%%%%%%%%%%%%%%%%%%%%%%%%%%%%%%%%%%%
\section{Method}
\label{sec:Method}
To identify the subset of 2RXS sources that are galaxy clusters detected due to their ICM X-ray signatures, we use the multi-component matched filter algorithm \citep[MCMF;][]{Klein18,Klein19}. We search for optical counterparts of 2RXS sources in the Legacy Survey, measuring redshifts and richnesses for each so that we can evaluate the probability that each is an X-ray selected galaxy cluster. This allows us to select a high-purity sample of candidate galaxy clusters from 2RXS or indeed to select several samples with different sizes and sample purities.  Thereafter, we combine multiple 2RXS detections of the same galaxy clusters, and finally we apply an additional point source rejection method, producing the final cluster catalog.

\begin{figure*}
\includegraphics[keepaspectratio=true,angle=0,width=0.485\linewidth]{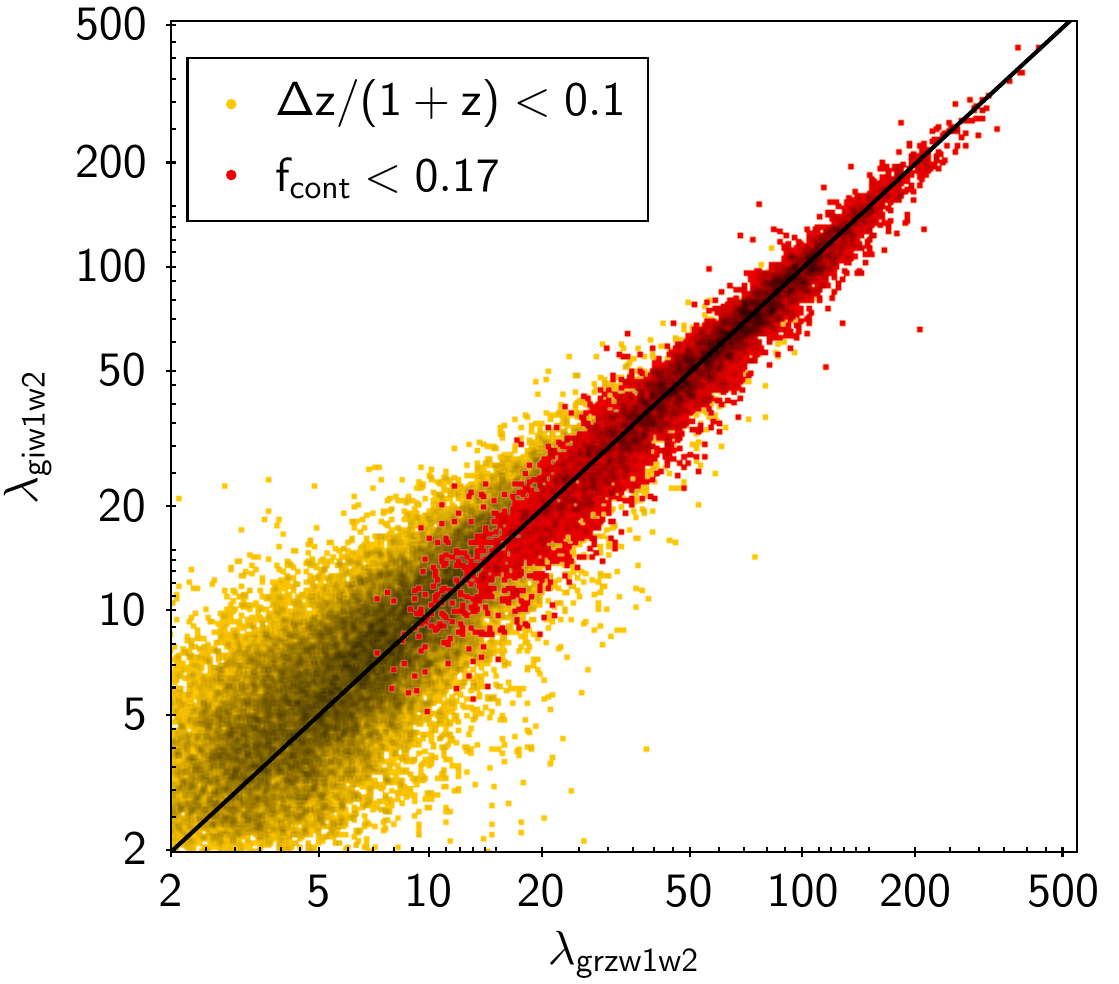}
\includegraphics[keepaspectratio=true,angle=0,width=0.50\linewidth]{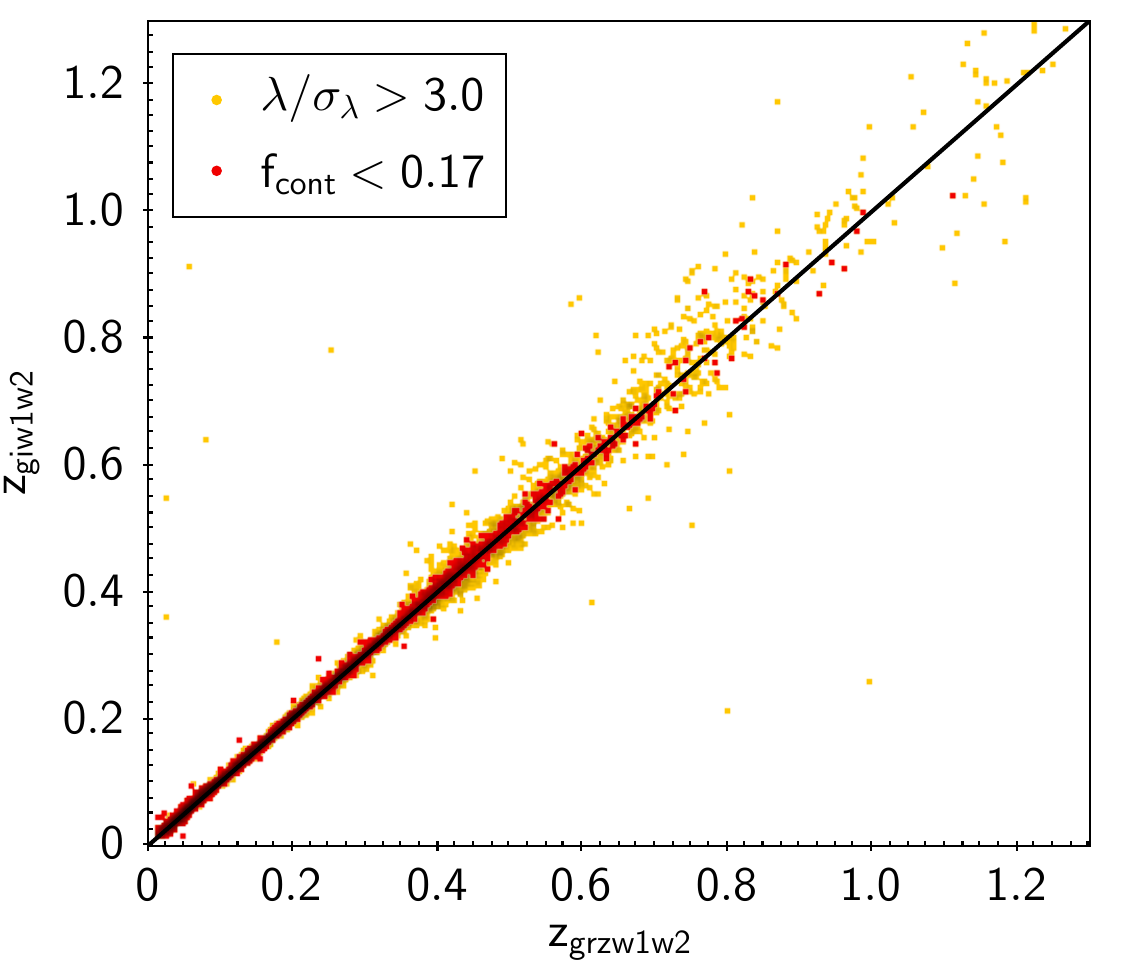}
\vskip-0.10in
\caption{Comparison of richnesses $\lambda$ and redshifts $z$ when using $grz$,$w1$,$w2$ to those for the same clusters when using $gi$,$w1$,$w2$ bands. In the left plot we show in yellow all sources with similar redshifts in both runs and in red the subset of clusters entering the final cluster catalog. The right plot shows all sources with richnesses greater than three times the richness uncertainty (yellow) and clusters making the selection into the final cluster catalog. The plots are heavily saturated containing $>33,000$ clusters on the left and $>11,000$ on the right.}
\label{fig:redshifts}
\end{figure*}

\subsection{Selecting a high-purity cluster sample}
\label{sec:MCMFselection}
MCMF was created for the identification of true clusters in ICM selected candidate catalogs. It was first applied to the same 2RXS catalog used in this work \citep{Klein18,Klein19} to create a galaxy cluster catalog over the DES area. Later it was also used to identify the first clusters identified by eROSITA over the eFEDS footprint \citep{Klein22} and to identify SZE selected clusters from the Planck survey \citep{Hernandez22}. We therefore present only a brief description in this work and refer the reader to the aforementioned publications for additional details.

The MCMF identification of clusters is based on the red sequence (RS) of cluster galaxies \citep{gladders00} and the weighted number-- called richness $\lambda$-- of excess RS galaxies within a certain magnitude and radial range around the X-ray position. The weights include a radial filter following a projected NFW profile \citep{navarro96} and a color-magnitude filter tuned to select red sequence galaxies. The color-magnitude filter has the form,
\begin{equation}
 w_i(z)=\frac{\prod\limits_{j=1}^{N}{G\left(c_{i,j}-\left<c(f,z)\right>_j,\sigma_{c_{i,j}}(f,z)\right)}}{N(\sigma_{c_{i,1}}
(f,z),\sigma_{c_{i,2}}(f,z),\sigma_{c_{i,3}}(f,z))}.
 \label{eq:color_weight}
\end{equation}
 Here $G\left(c_{i,j}-\left<c(f,z)\right>_j,\sigma_{c_{i,j}}(f,z)\right)$ is the value of the normalized Gaussian function at a color offset between observed color $j$ and predicted RS color given observed reference band magnitude $f$ of source $i$ and assumed redshift $z$. 

 In the current analysis, we run MCMF in two different set of bands. The $grz$-mode, uses the combinations $c_{j}$ of $g-r$, $r-z$, $z-w1$ and $w1-w2$. In this mode we use the $z$-band magnitude as the magnitude reference band $f$.
 To maximize the footprint we also run MCMF in the $gi$-mode which uses $g-i$, $i-w1$ and $w1-w2$. Here we use the $i$-band for the reference band $f$. 
 The standard deviation of the Gaussian weight function is the combination of intrinsic and measured scatter as $\sigma_{c_{i,j}}(f,z)=\sqrt{\sigma^2_\mathrm{mcor}(f,z) + \sigma^2_\mathrm{meas,i}}.$

Given the additional redshift information provided by adding Mid-IR WISE data to the optical set of colors we expand the redshift range compared to previous runs from $z=1.3$ to $z=1.5$ and for each sample we scan through 300 redshift bins calculating the richness. For each redshift bin an aperture corresponding approximately to $r_{500}$ is estimated, based on the X-ray count rate, the redshift of the bin and an observable-mass scaling relation. 

Following previous MCMF analyses \citep{Klein18,Klein19}, we convert the observed X-ray count rate to an estimate of the X-ray luminosity assuming an APEC plasma model \citep{Smith2001} of fixed metallicity (0.4~solar) and temperature (5~keV). The conversion factor to luminosity is then derived for the Galactic neutral hydrogen column density and redshift of each source, given the ROSAT instrumental response.
We adopt the same luminosity-mass scaling relation as in the previous MCMF work over the DES footprint,
\begin{eqnarray}
\label{eq:xray_type_II}
L_{500,0.5-2.0 \mathrm{~keV}} = A_\mathrm{X}\left( \frac{M_{500}}{M_\mathrm{piv}} \right)^{B_\mathrm{X}} 
\left(\frac{E(z)}{E(z_\mathrm{piv})}\right)^2
\left( \frac{1+z}{1+z_\mathrm{piv}} \right)^{\gamma_\mathrm{X}},
\end{eqnarray}
where $A_\mathrm{X}$, $B_\mathrm{X}$ and $\gamma_\mathrm{X}$ have best values of $4.15\times10^{44}$~erg s$^{-1}$, 1.91 and 0.252 respectively. The redshift pivot is 0.45 and the mass pivot is given as $6.35\times10^{14} M_\odot$ \citep{bulbul19} .

The galaxies contributing to the richness measurement $\lambda$ are not only limited to be within $r_{500}$ but also within a certain luminosity range. Compared to our previous work we expand this range to be $m^*-3$ to $m^*+2$, where the characteristic magnitude $m^*$ is the same as in our previous work and is based on a star formation model with an exponentially decaying starburst at a redshift z = 3 with a Chabrier initial mass function
and a decay time of 0.4 Gyr \citep{BC03}. We therefore now consider galaxies 0.75~mag fainter than before, which leads to a $\sim1.5$ times increase in $\lambda$ compared to runs with the previous RS galaxy luminosity cut. We trace the local imaging depth at the location of a cluster candidate and account for missing sources in cases where the local imaging depth does not reach $m^*+2$ (for further details see Sec. 3.5 in \citet{Klein19}).

The distribution of richness given redshift is then searched for peaks and then fitted by so-called peak profiles, which are derived from stacks of clusters with spectroscopic redshifts \citep[see][for examples]{Klein19,Klein22}. With this approach we directly perform a calibration against spectroscopic redshifts and include other effects such as the contribution of blue cluster members or the evolution of the used aperture ($r_{500}$) as function of redshift into the profile shape.

\subsubsection{Identifying and removing the contamination}

Separating non-cluster sources from real X-ray clusters requires that we estimate the probability that a source with measured redshift and richness is a contaminating source. A contaminating source would be an AGN, star or 2RXS noise fluctuation that happens to lie along the line of sight toward a physically unassociated optical system.
With the MCMF algorithm, we use the differences between the richness distributions as a function of redshift $f(\lambda,z_i)$ toward contaminants and toward galaxy clusters, to then assign an estimate that each matched pair is a random superposition of X-ray and optical source rather than a true cluster. The cluster candidates with the highest probability of being contamination are excluded from the catalog.

To estimate the richness distribution of the contaminants $f(\lambda,z_i)$ one wants to apply MCMF to a catalog that represents as much as possible the characteristics of these contaminants. In the case of 2RXS the vast majority of the sample ($\sim85\%$) is either AGN, star or noise fluctuation, and we therefore make use of the 2RXS catalog itself. In previous work on the MARD-Y3 sample \citep{Klein19}, randomly redistributed 2RXS positions were used to characterize the richness distribution of the contaminants. 
In fact, it is important to trace possible changes in the properties of the contaminants as a function of survey characteristics such as exposure time or location on the sky or with respect to the galactic plane. Therefore, in the current analysis we systematically shift the 2RXS source positions along ecliptic latitude, because this follows the scan direction of the RASS survey. We create four shifted versions of the 2RXS catalog shifted by plus and minus one and two degrees. This largely preserves characteristics such as the exposure time distribution, source density and flux distributions. We apply MCMF to those shifted catalogs after removing any shifted locations that by chance correspond to the locations of real 2RXS sources in positional and redshift space, because the goal is to measure the richness distributions of non-2RXS selected sources and in particular to avoid biasing of these richness distributions by galaxy clusters.
We refer to the resulting richness distribution as $f_\mathrm{rand}(\lambda,z_i)$. 

Similarly, we measure the richness distribution of the 2RXS candidates $f_\mathrm{obs}(\lambda,z_i)$.
With the set of richness distributions from the 2RXS catalog and the shifted catalogs we calculate for each candidate $i$ the contamination estimator $f_\mathrm{cont,i}$, which is defined as
\begin{equation}
f_{\mathrm{cont},i}=\frac{\int_{\lambda_i}^{\infty} f_\mathrm{rand}(\lambda,z_i) d\lambda}{\int_{\lambda_i}^{\infty}
f_\mathrm{obs}(\lambda,z_i) d\lambda},
\end{equation}
where $z_i$ and $\lambda_i$ are the redshift and richness of the cluster candidate, $f_\mathrm{obs}(\lambda,z_i)$ is the richness distribution of 2RXS candidates and $f_\mathrm{rand,z}(\lambda,z_i)$ is the richness distribution extracted from the shifted catalogs.  The integrands are evaluated at the candidate redshift $z_i$ and the integral is carried out from the candidate richness $\lambda_i$. 

The first step we take in defining a cluster sample is to adopt an \fcont\ threshold $f_\mathrm{cont}^\mathrm{cut}$.  Such a sample has an associated $\lambda_\mathrm{min}(z)$ that marks the minimal richness a cluster at redshift $z$ must have to be included in the sample.  For cluster samples that extend to very low mass ($\sim10^{14}M_\odot$), this \fcont\ selection introduces incompleteness, which can be accounted for using the $\lambda$-mass relation describing the sample.

By design, a cluster sample selected with $f_{\mathrm{cont},i}<f_\mathrm{cont}^\mathrm{cut}$ has its contamination reduced by the factor $f_\mathrm{cont}^\mathrm{cut}$ as long as the shifted catalogs (i.e., random positions) produce richness distributions $f_\mathrm{rand}(\lambda,z)$ that are representative of the contaminating sources. 
In this work we may expect some small difference between the richness distributions $f_\mathrm{rand}(\lambda,z)$ we use and that of the contaminants. Only $\sim35$\% of the contaminants are expected to be noise fluctuations and another $\sim15$\% are stellar X-ray sources.  Neither of these source types are correlated in any way with the passive galaxy population in the Universe.
However, the remaining ($\sim65$\%) of the contaminants are X-ray AGN, which are hosted by galaxies and therefore trace the large scale structure as do the cluster passive galaxies. Therefore, in Section~\ref{sec:catpurity} we carry out a validation by measuring the contamination of the catalog as function of $f_\mathrm{cont}^\mathrm{cut}$.

\subsubsection{Constraining the impact of different band combinations}
As described in Section~\ref{sec:datals}, the Legacy Survey can be broadly divided into three parts, Legacy Survey North, Legacy Survey South $grz$ and Legacy Survey South $gi$. Using overlapping regions between the north and south parts, we calibrate the galaxy photometry of the north part to the system in the south. We further create individual peak profiles for each of the three MCMF runs based on spectroscopic clusters in those subsamples.

The photometric redshift and richness measurements for overlapping sources of the $grz$ runs of the North and South surveys are in decent agreement with  scatter in redshift of $\sigma_{z/(1+z)}=0.003$ and $\sigma_{\ln(\lambda_\mathrm{North}/\lambda_\mathrm{South})}=0.12$. We further compared \fcont\ measurements between both survey patches.
We also found agreement too, which means that also the distribution of the richness estimates around randoms is comparable. We therefore merge both of the $grz$ based MCMF runs to one big catalog and calculate \fcont\ from the merged 2RXS and shifted catalogs.

We additionally did an MCMF run using the full set of $griz$ bands. The performance improvements in photometric redshifts were minor, not justifying yet another subdivision of the survey.

The sole reason for running MCMF in $gi$-mode is to increase the footprint solid angle, in particular to fill a $\sim$1000~deg$^2$ hole in the eastern galactic hemisphere with low $rz$-band coverage. There is a large overlap between the MCMF run in $grz$-mode and that using the $gi$-mode.

In Figure~\ref{fig:redshifts} we show the redshift and richness comparison between the two runs. We do not find any significant redshift bias between both systems and an outlier fraction of 0.2\% (0.8\%), for candidates with $\lambda/\sigma_\lambda>3$ and defining offsets of $\Delta z/(1+z)>0.1 (0.05)$ as outliers.
We find a richness trend of $\lambda_{grzw1w2}=1.1 * \lambda_{giw1w2}$ 
that we correct before we include sources from $gi$-mode into the main sample.

Given that the sky area and the number of clusters identified in the $gi$-mode run are small, and given that the $gi$-mode performance is similar to that of the $grz$-mode, we adopt the $grz$-mode mapping of richness and redshift to \fcont. This means that for a given \fcont\ based selection, the same redshift dependent richness cut applies to all survey regions, independent of whether it lies in the Legacy Survey North, the South or the region with the $gi$-mode measurements. Note that we provide footprint related flags in the catalog to allow users to compare different parts of the survey.

\begin{figure}
\includegraphics[keepaspectratio=true,angle=270,width=1\linewidth]{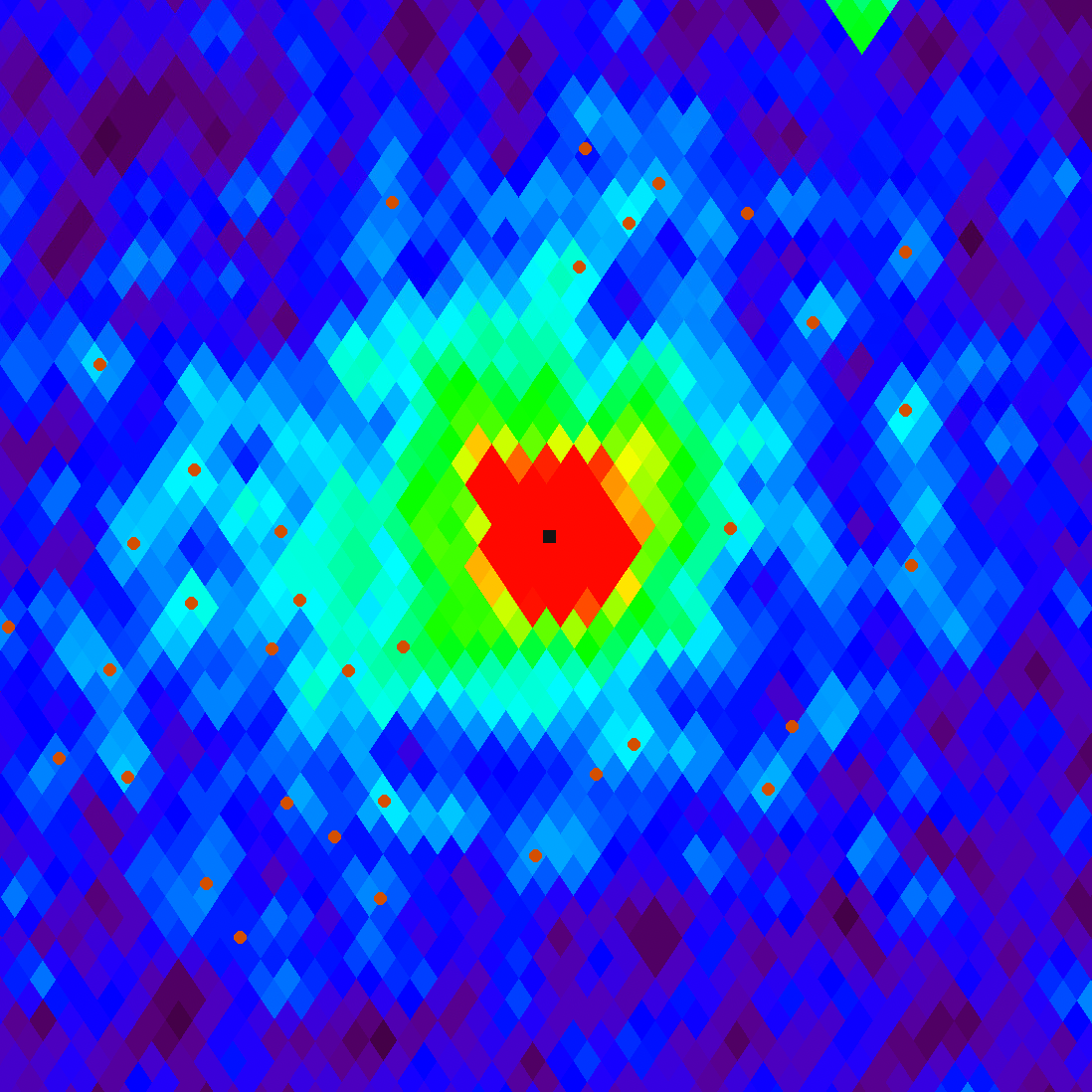}
\vskip-0.0in
\caption{Smoothed RASS X-ray count rate map of a $40'\times40'$ region centered on the Centaurus cluster (ACO~3526). There are 36 2RXS sources in this field. Only one source (black square) survives the \fcont selection and the rejection of multiple detections of the same source.  Orange circles mark those sources identified as multiple detections.}
\label{fig:Centaurus}
\end{figure}

\subsection{Rejecting multiple detections of the same cluster}
\label{sec:rejectingmultiples}
One known problem with the 2RXS catalog and other X-ray catalogs based on the same detection pipeline is that bright, extended X-ray sources are multiply detected and listed in the catalog. 2RXS provides an optical screening flag \citep[S\_FLAG;][]{Boller16}{}{} to partially address this problem, but we have found that this flagging is neither complete nor is it always correct. We therefore make use of the optical cluster centres to identify multiple detections, which we then remove from the catalog.  Important in this process is to tune the removal so that resolved galaxy cluster mergers and pairs remain in the catalog.

As described in previous MCMF analyses \citep{Klein18,Klein19,Klein22}, each cluster has two possible centres: 1) the BCG position and 2) the position of the peak of the galaxy density map.
The galaxy density map is created using the same color weights as for the richness measurements, and therefore represents the density of red sequence like galaxies at the cluster redshift. The density peaks are searched over the full size of the map ($\sim5$~Mpc) and the nearest peak is recorded as the galaxy density based centre.
The BCG is identified as the brightest galaxy within 1.5 Mpc, $3\sigma$ from the mean red sequence color and brighter than $m^*-1$. If no source is found we expand the search to $4\sigma$ and $m^*-0.5$. If still no source is found, we consider the BCG search to be unsuccessful. We consider the galaxy density centre to be the more robust estimate, while the BCG position is the more accurate, if correctly identified. We consider the BCG to be correctly identified if it agrees within 250~kpc with either the X-ray position or the galaxy density centre. In 81\% of the cases we consider the BCG to be correctly identified, for the remaining cases we use the galaxy density centre.

To flag multiple detections of the same cluster we first flag all 2RXS sources where the optical centre has a nearer match to another 2RXS source whose redshift differs by $\Delta z\le0.05$.  
We then run through the list of flagged sources and check if this criterion results in flagging all sources of a given system. We then include in the catalog the system with highest count rate and exclude the others.
Finally, we merge the measured count rates of all multiple detections closer than 1 Mpc that are not flagged as likely point-like sources from the dedicated point-like follow-up discussed in the following section.

\subsection{Identifying residual point source contamination}
\label{sec:pointsourcecontamination}

The high-purity cluster samples created from an MCMF run still have residual point source contamination-- these are essentially all random superpositions of X-ray point sources with red sequence optical systems.
The unresolved X-ray sources have been studied extensively to 
identify the 2RXS AGN and stars and to assign the best associated optical or infrared counterparts. In the following subsections we summarize this work, because we use the results to devise a method for removing many of the contaminating AGN and stars that initially make it into our cluster sample.

\subsubsection{2RXS detected Stars}\label{sec:stellar}

Young and fast rotating stars show a high ratio of X-ray to bolometric luminosity of up to $L_\mathrm{X}/L_\mathrm{bol}=10^{-3}$, which seems to be a saturation limit \citep{Vilhu84,Wright11}. For slow rotating old stars the ratio of luminosities can be as low as $L_\mathrm{X}/L_\mathrm{bol}=10^{-8}$ \citep{gudel04,testa15}. Given the 2RXS flux limit, these luminosity ratios imply 
that 2RXS stellar sources should be brighter than 15~mag and within a distances of $\sim$750~kpc \citep{freund22}; therefore, they should be included in the GAIA catalogs \citep{gaia,gaiaedr3}. 
In fact, the limitation in identifying optical counterparts for 2RXS stars is that some of those stars will be too bright for GAIA, making it sensible to augment the GAIA catalog with the Tycho-2 catalog \citep{tycho2}. 

In their analysis of 2RXS stars, \citet{freund22} 
calculated matching probabilities using positional offsets. 
The quantity $p_{ij}$ gives the probability that a given stellar source is the right counterpart to the 2RXS source and $p_\mathrm{stellar}$ is the probability that any of the given stellar counterparts is a match to the 2RXS source and therefore the probability that the 2RXS source is a star. 
Due to the poor angular resolution of RASS, there are often many possible optical counterparts, and thus two further observables-- the X--ray to Gaia flux ratio and the stellar distance-- were used to help improve the counterpart selection.
Because source densities characteristics of true and contaminating sources change with sky position, the whole analysis and calibration is performed independently on multiple patches on the sky, depending on galactic coordinates. With improved selection algorithm, it is possible to create a $\sim 93\%$ pure and complete stellar sample using $p_\mathrm{stellar}>0.51$ and $p_\mathrm{ij}>0.5$ \citep{freund22}. The fraction of stars in 2RXS depends on galactic latitude \citep[see Figure 5;][]{freund22}, but over the region of interest of our study (excluding galactic plane), the 
stars constitute only $10-15$\% of the 2RXS sources.

\begin{figure}
\includegraphics[keepaspectratio=true,width=0.99\linewidth]{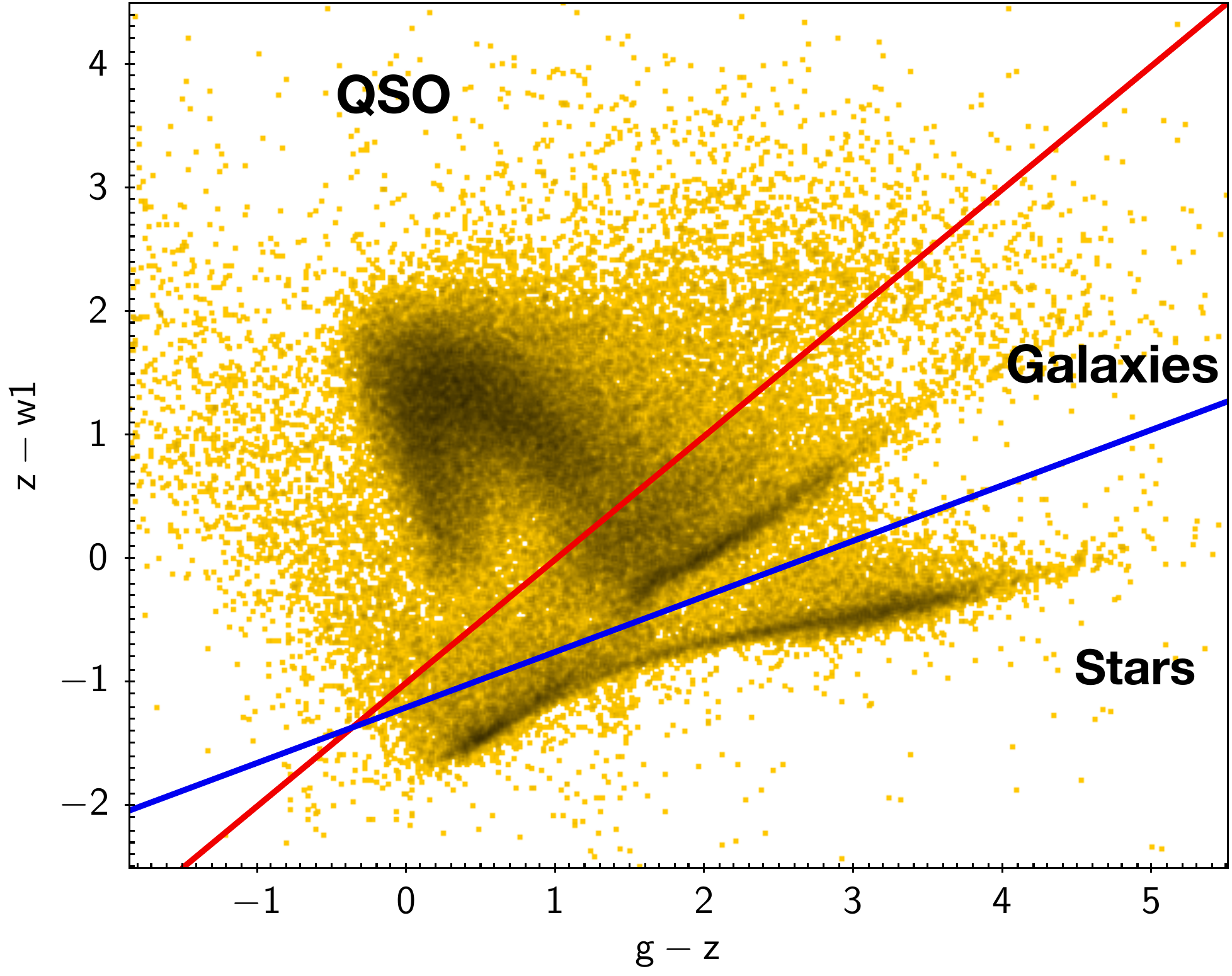}
\caption{Color distribution of NWAY selected counterparts of 2RXS sources.  Several high-density regions are visible, which can be separated by two lines. Sources above the red line are predominantly QSO. The high-density region below the blue line belongs to stellar sources. Sources in between the two lines correspond to galaxies, where red sequence galaxies build a dense stripe starting at $g-z\approx 1.5$}
\label{fig:PScolors}
\end{figure}

\subsubsection{2RXS counterparts}\label{sec:NWAY}
A dedicated search for AllWISE \citep{Cutri13} counterparts to 2RXS sources \citep{Salvato18} 
focused on the identification of extragalactic point sources, limiting the analysis to $\vert b\vert > 15$
and excluding regions of radius $6^\circ$ and $3^\circ$ around the Large and Small Magellanic Clouds (similar to the region in 
Figure~\ref{fig:footprint}).

To identify counterparts, a Bayesian statistics based algorithm called NWAY was used, that adds priors from counterpart magnitudes and colors to the typical counterpart search based on position and source density \citep{Salvato18}. To inform the color-magnitude prior, 2,349 secure counterparts from 3XMM-DR5 \citep{ROSEN16} were used. The application of NWAY to 2RXS resulted in at least one ALLWISE candidate for 99.9\% of the 2RXS catalog in the previously defined footprint and within the a maximum offset of 2~arcminutes.

Key NWAY measurements used to identify good counterparts are $p\_any$ and $p\_i$. The estimator $p\_any$ provides the probability that any of the considered ALLWISE counterparts are the correct 2RXS counterpart.
%and therefore is likely a real source. 
The second estimator $p\_i$ provides the probability of a given ALLWISE candidate to be the correct counterpart. High values of $p\_i$ therefore indicate clear one-to-one matches between 2RXS and ALLWISE. The analysis found $\sim59\%$ of the 2RXS sources with $p\_any>0.5$, while only 5\% of randomized 2RXS positions show $p\_any>0.5$ and $p\_i>0.8$. Given that 2RXS is expected to have 30\% spurious sources, the expected contamination by spurious sources after applying this cut to 2RXS is $\leq 2\%$. Based on the same assumption, this cut provides counterparts to $84\%$ of real sources in 2RXS. Considering multiple detections caused by extended sources ($\sim 4.5\%$) the fraction of identified sources increases to 90\%. Given the optimization to point-like sources and the presence of clusters with significant positional offsets, this suggests that the completeness of identifying AGN in 2RXS is likely significantly higher than 90\% \citep{Salvato18}.

\subsubsection{Photometric properties of 2RXS counterparts}
Thanks to the much better positional accuracy from the WISE counterpart, compared to 2RXS, we can simply match the best ALLWISE counterparts to the LS~DR10 data set using a 1.5 arcsec maximum offset. With that, we have $g, r, z$ and forced WISE $w1,w2$ photometry for the point-like counterparts in our footprint. As the NWAY catalog contains counterparts to various source types, including the BCG of galaxy clusters, the color information available allows us to split between source types. In Figure~\ref{fig:PScolors} we show the color distribution of NWAY counterparts in $z-w1$ vs $g-z$ color. Visible are three over dense regions, that can be roughly separated by the two lines drawn in the plot. The 
%bluest (What?)
over density below the blue line ($z-w1=(g-z)\times 0.4 - 1.2$) is the stellar locus, which is dominated by stars. The over density above the red line ($z-w1=(g-z) - 1$) is dominated by AGN, mostly by QSOs. The over density in between the lines is dominated by passive, red-sequence like galaxies.
In our subsequent analysis of contamination by non-cluster sources in Section~\ref{sec:catpurity} we make use of the LS~DR10 photometry to define a clean AGN subsample.

%%%%%%%%%%%%%%%%%%%%%%%%%%%%%%%%%%%%%%%%%%%%%%%%%%%%%
%%%% catalog creation
%%%%%%%%%%%%%%%%%%%%%%%%%%%%%%%%%%%%%%%%%%%%%%%%%%%%%
\section{RASS-MCMF cluster catalog}
\label{sec:Catalog}
There are three key criteria that affect the usefulness of a cluster catalog: sample purity, sample size and the difficulty of modelling 
the selection function. Typically there is a trade off between sample size and purity, because methods to remove contamination from the sample often also remove some real clusters. In addition, more complicated cleaning methods will likely minimize this loss but may make the sample selection more challenging to model.  That can then impact one's ability to do cosmological studies with the sample.

To make an educated choice of the final sample definition, we first characterise the sample purity and estimate the impact of the additional point source rejection step.
%and the potential impact on the selection function. 
We then define the multi-component matched filter RASS cluster catalog confirmed with DESI Legacy Surveys (RASS-MCMF) sample in Section~\ref{sec:CatalogDefinition}, presenting the characteristics of the 90\% purity RASS-MCMF catalog together with two subsets of the catalog that have 95\% and 99\% purity.  In Section~\ref{sec:spectroscopicredshifts} we describe the cross-matching required to assign spectroscopic redshifts to over half the RASS-MCMF clusters.  Finally, in Section~\ref{sec:properties} we discuss some properties of the sample, including mass estimates, redshifts and richnesses.

\subsection{Measuring catalog contamination}\label{sec:catpurity}
To measure the level of contamination by non-cluster sources (AGN, stars, noise fluctuations) we follow previous MCMF analyses \citep{Klein19,Klein22} and model the distribution of the contamination and the clusters in an observable-observable space $\log (\lambda/M_X)$. As noted previously, $M_X$ is an X-ray based mass estimate that uses the object X-ray flux and redshift (see equation~\ref{eq:xray_type_II}).  

\begin{figure}
\includegraphics[keepaspectratio=true,width=1\linewidth]{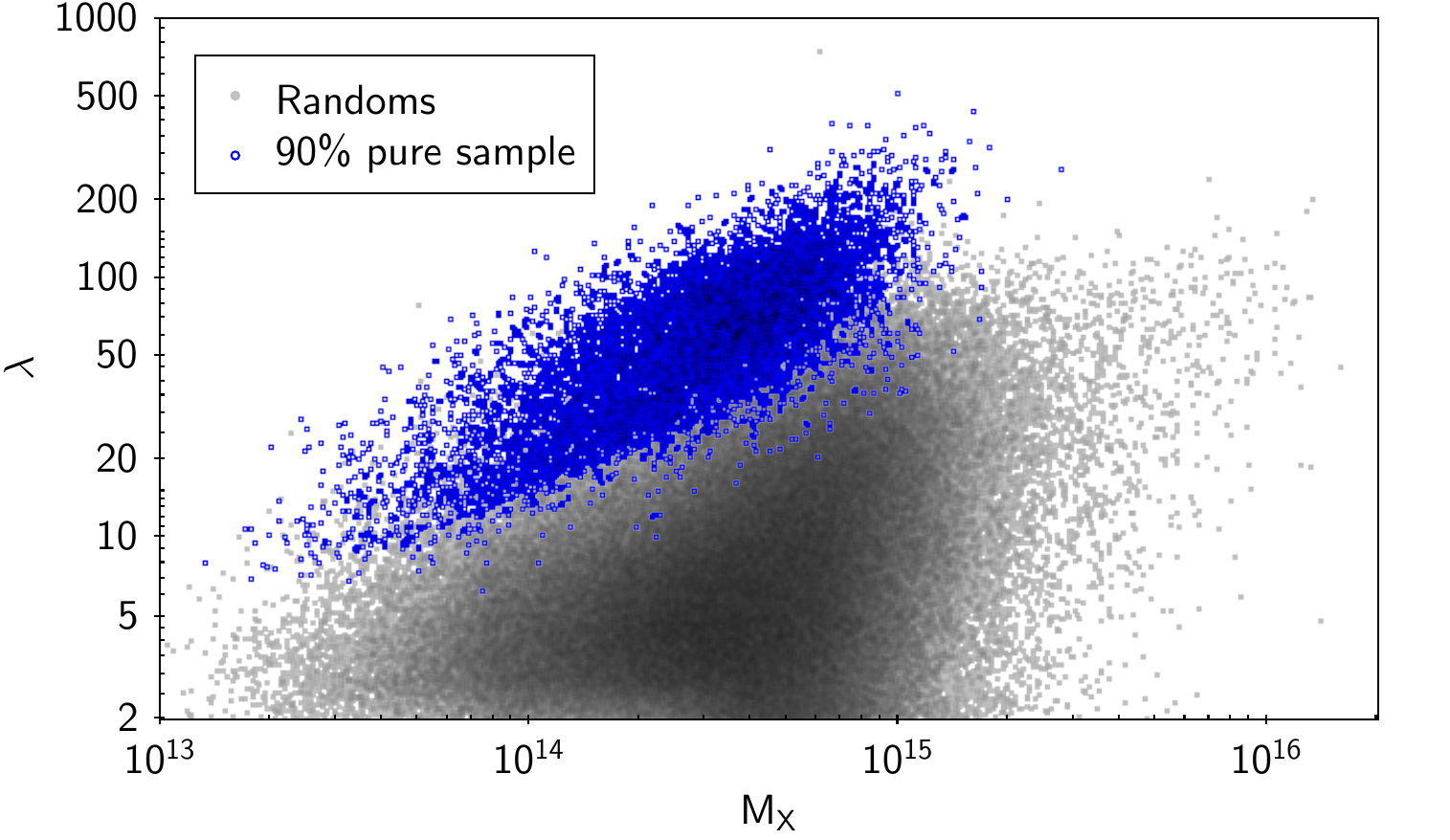}
\caption{Distribution of RASS-MCMF clusters (blue) and non-clusters selected along random lines of sight (gray) in $\lambda$ versus $M_X$, where $M_X$ is an X-ray mass estimate that assumes the source is a cluster at the redshift of the best optical counterpart. Real clusters scatter about the $\lambda$-$M_X$ relation while non-clusters extracted along random lines of sight are predominantly distributed below that relation. }
\label{fig:LambdaMX}
\end{figure}

Figure~\ref{fig:LambdaMX} shows the distribution of a clean sample, defined using an \fcont\ threshold as desribed in the following subsection and that of the shifted 2RXS catalogs (i.e., random sky locations within the survey). Real clusters follow the power-law observable-observable relation, which for $\lambda-M_X$ is approximately a relation with slope of one, and exhibit considerable scatter.
In contrast, the distribution of the random sky locations in this space lies significantly lower than the clusters.  Essentially, the clusters are the 2RXS sources with the highest richnesses.  The density of the contamination in this space peaks well below the clusters, but at a given mass the tails of the cluster and contamination distributions overlap somewhat in richness. 

\begin{figure}
\includegraphics[keepaspectratio=true,width=1\linewidth]{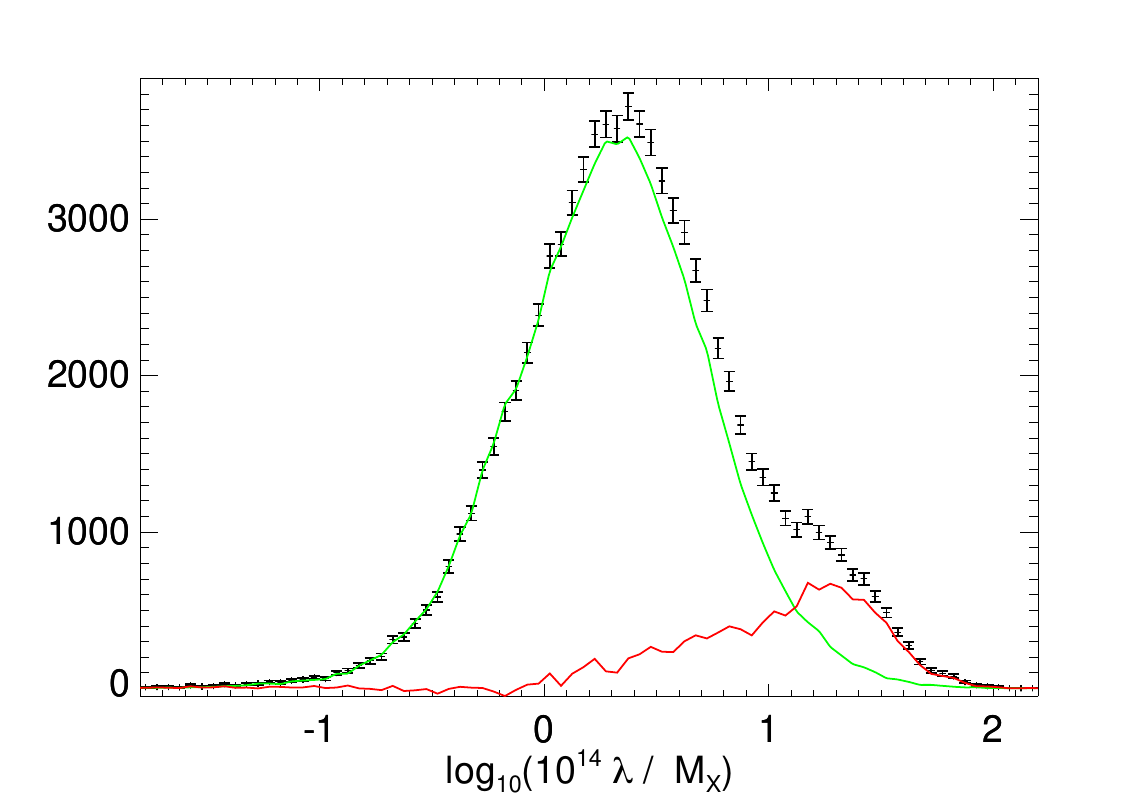}
\includegraphics[keepaspectratio=true,width=1\linewidth]{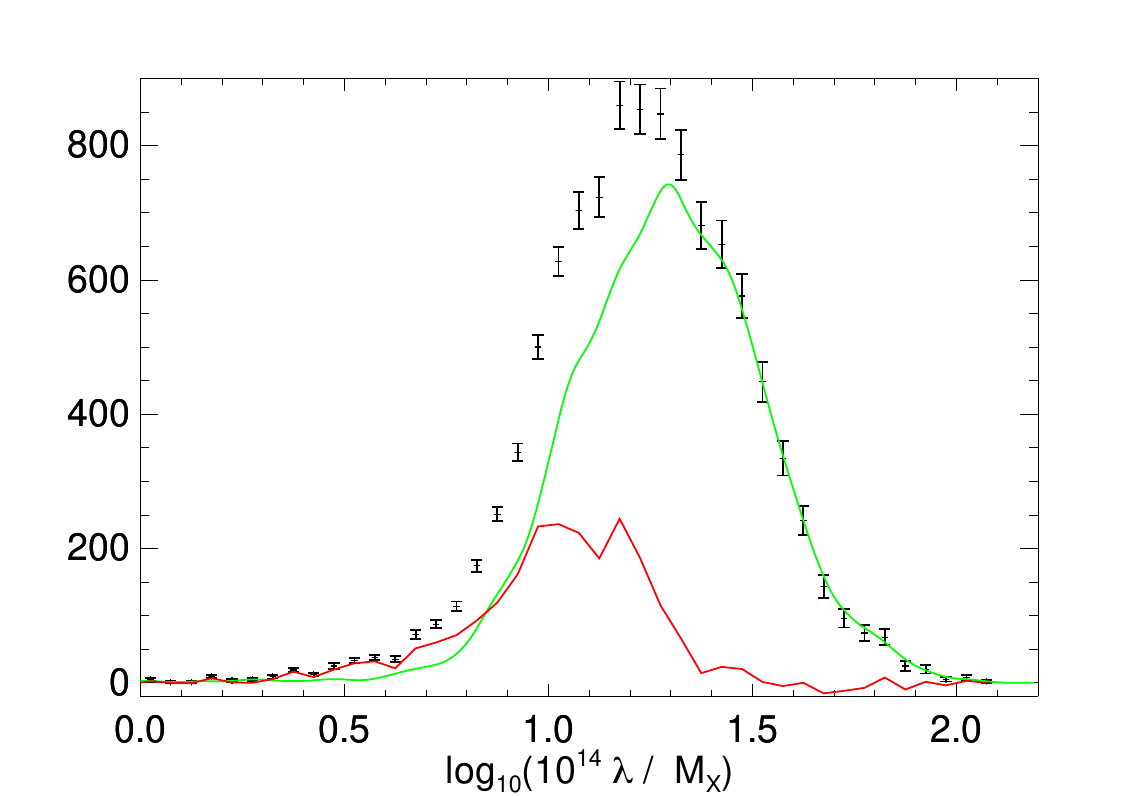}
\caption{Histogram (top) of the richness over X-ray mass ratio in $\logTen(10^{14}\lambda/M_X)$ for all 2RXS sources in the optical footprint, excluding multiple detections of the same cluster. Green line show the model of the contaminant population; the red line shows the residual between the contaminant model and total distribution, which is the estimate for the cluster population. Similar plot (bottom) but for the subsample with \fcont < 0.2. Here a cluster model (green) is fit to the data and the residual (red) is showing the estimated distribution of non-clusters.}
\label{fig:Contamodelfc02}
\end{figure}

In the top panel of Figure~\ref{fig:Contamodelfc02} we show the distribution in $\log(10^{14}\lambda/M_X)$ for 2RXS sources in Legacy Survey DR10 (black data points), excluding only the multiple detections of the same clusters.  Given the relative behavior of clusters and contamination in Figure~\ref{fig:LambdaMX}, it is clear that clusters prefer higher $\log(10^{14}\lambda/M_X)$.  

For a measurement of the contamination fraction of the full candidate list, we model the $\log(10^{14}\lambda/M_X)$ distribution with a contamination model only (green line; described below), limiting the fit region to low-enough $\log(10^{14}\lambda/M_X)$ values that contamination by real clusters is minimal.
We do not simultaneously fit for a cluster model, because clusters compose only a fraction of the 2RXS catalog, and the statistics are not adequate to produce a good model.

For the contamination model we consider three different populations: noise fluctuations, AGN and stellar sources. As already mentioned, the AGN are hosted in galaxies that trace the large scale structure, whereas the noise fluctuations and bright stars are uncorrelated with it. 

The 2RXS catalog paper provides estimates for the fraction of noise fluctuations as a function of existence likelihood. To create a model for noise fluctuations, we select a subset of the sources along random lines of sight that follows this expected distribution in existence likelihood \citep[see][]{Boller16}.  
The X-ray AGN model is based on sources directly selected from 2RXS sources using NWAY selection thresholds as described in Section~\ref{sec:NWAY} and a color cut shown in Figure~\ref{fig:PScolors} that excludes passive galaxies and stars. While the NWAY thresholds ensure a sample of $\sim$ 98\% real sources, the color cut excludes non-AGN like sources from the sample. 
For the stellar subsample we use the selection described in Section~\ref{sec:stellar}, yielding 93\% purity and completeness. 

We then combine all three models to build a composite contamination model, assuming the constituent fractions are 30\%, 43\% and 12\%, which leaves space for $\sim15$\% clusters. We do not attempt to fit for the relative contributions of AGN, stars and spurious sources, because all three models of contaminants are similar enough that they cannot be independently constrained with the given data.
When restricting the fit to $\log(10^{14}\lambda/M_X)<0$ (see Figure~\ref{fig:Contamodelfc02} top) we find a contamination of $87\pm2$\%, which provides an estimate of the contaminant population and expected number of clusters in the sample. Previous work on RASS data showed that between 16\% \citep{hasinger21} and 20.5\%  \citep{boehringer13} of the RASS sources should be galaxy clusters. Accounting for the fact that $\sim 30\%$ of 2RXS siources are noise fluctuations, we obtain the fraction of clusters among the real X-ray sources to be $18.5\pm3.0\%$. This is in good agreement with previous estimates and suggests that 2RXS should contain $>10,000$ galaxy clusters in the extragalactic sky.

To enable a test of the true contamination fraction in \fcont\ selected samples such as RASS-MCMF, we must estimate the level of contamination of a cluster dominated subsample. 
To do this we first exclude all likely stellar and AGN sources (using the same selection methods described above), and reduce the spurious sources by increasing the existence likelihood cut to 8.08, corresponding to a reduction from 30\% to 10\% \citep[according to][]{Boller16}. The stellar and AGN rejection excludes 93-98\% of real point sources leaving a sample with only $\sim$15\% residual contamination. Creating a cluster sample from this cleaned catalogue using a threshold \fcont<0.3 would then further reduce the contamination by a factor of three, creating a $\sim$95\% pure cluster sample.  At even smaller \fcont\ thresholds, the contamination in the cluster sample quickly becomes irrelevant.  

We use this clean cluster sample to create a cluster model at the high (cluster dominated) end of the $\log(10^{14}\lambda/M_X)$ distribution, which will enable us to estimate the number of real clusters in any subsample selected with a particular threshold \fcont.

\begin{figure}
\includegraphics[keepaspectratio=true,width=1\linewidth]{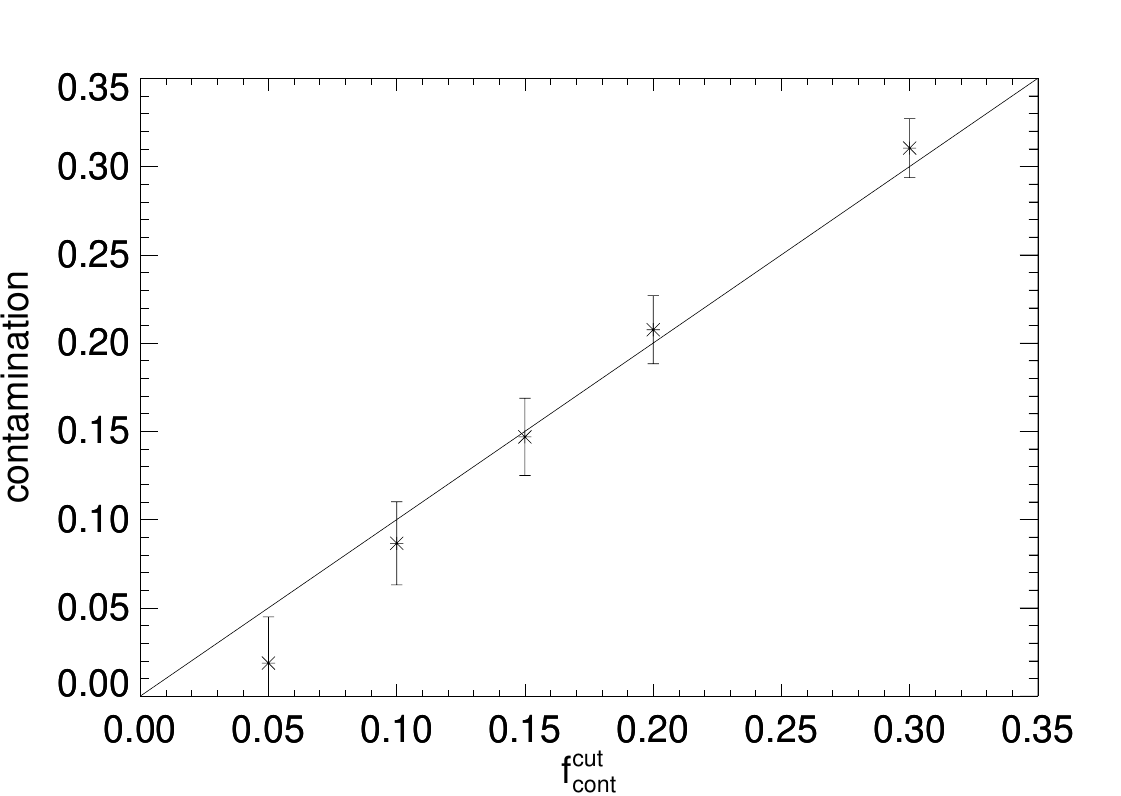}
\caption{Estimated contamination from the cluster model fit to the $\lambda/M_{500}$ distribution (see Figure~\ref{fig:Contamodelfc02}) versus the \fcont\ selection threshold $f_\mathrm{cont}^\mathrm{cut}$.  The line marks the expected contamination of each subsample, given the initial contamination measured to be $87\pm2$\%.  The measurements are in good agreement with expectations.}
\label{fig:ContavsFcont}
\end{figure}

We measure the contamination fraction for multiple \fcont\ thresholds and show them in Figure~\ref{fig:ContavsFcont}. If we would assume the real contamination is well described by the richness distribution extracted using the shifted 2RXS catalogs, we would expect the final contamination of the sample to be the \fcont\ threshold times the initial contamination of the candidate list ($\sim87\pm2$\%). 
The measurements shown in Figure~\ref{fig:ContavsFcont} suggest
that the actual contamination within \fcont\ selected samples is at the expected level for \fcont\ thresholds between 0.05 and 0.3.  In other words, there is no evidence that richness distributions $f_\mathrm{rand}(\lambda,z)$ derived from the shifted 2RXS catalogs (corresponding therefore to random lines of sight) differ from the true richness distributions of the non-cluster contaminants in the 2RXS catalog.

\subsection{Additional point source removal}
\label{sec:additionalPSremoval}

From the cluster fits to the observed $\logTen(10^{14}\lambda/M_X)$ distribution, we know the amount
of contamination for any given threshold in \fcont, and we know that contamination predominantly lies at low $\logTen(10^{14}\lambda/M_X)$ values (see bottom panel of Figure~\ref{fig:Contamodelfc02}). The distribution is dominated by the contamination for $\logTen(10^{14}\lambda/M_X)<1.0$, indicating that excluding identifiable stars and AGN with QSO colors that exhibit low $\logTen(10^{14}\lambda/M_X)$ would be effective at reducing sample contamination while having only a minor impact on the real cluster content of the sample.  

We test this explicitly by examining the multiband optical images of low $\logTen(10^{14}\lambda/M_X)$ sources, finding the first clear cluster cases at $\logTen(10^{14}\lambda/M_X)\sim1.1$.
We therefore set $\logTen(10^{14}\lambda/M_X)\approx1.1$ as the upper limit for point source rejection and exclude all those NWAY identified sources with QSO colors and all Gaia selected stars from the sample in this region.

We emphasize that some sources identified as AGN or stars could be associated with clusters. 
With the reported purities of the stellar sample (93\%) and the NWAY counterparts (98\%) (see discussion of stellar and AGN selection in Section~\ref{sec:pointsourcecontamination}), we estimate that $\sim$95\% of the excluded sources are true contamination, with the remainder being clusters.
Using this information, we can estimate the fraction of lost clusters coming from direct point source exclusion.  For the RASS-MCMF 90\%, 95\% and 99\% purity cluster samples presented below, this introduces an effective cluster incompleteness of 
0.7\%, 0.3\% and 0.2\%, respectively.

\begin{figure*}
\includegraphics[keepaspectratio=true,width=0.50\linewidth]{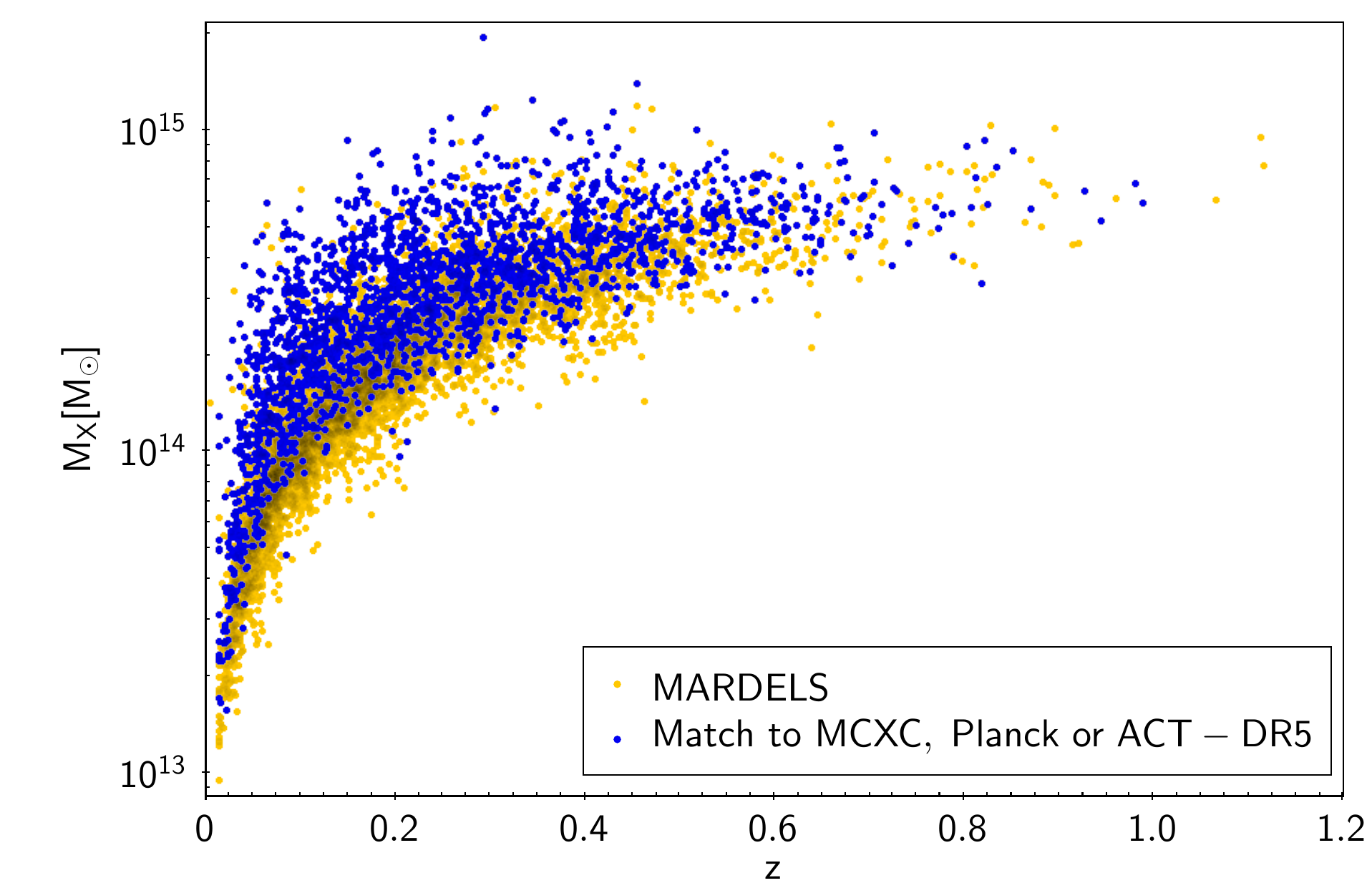}
\includegraphics[keepaspectratio=true,width=0.49\linewidth]{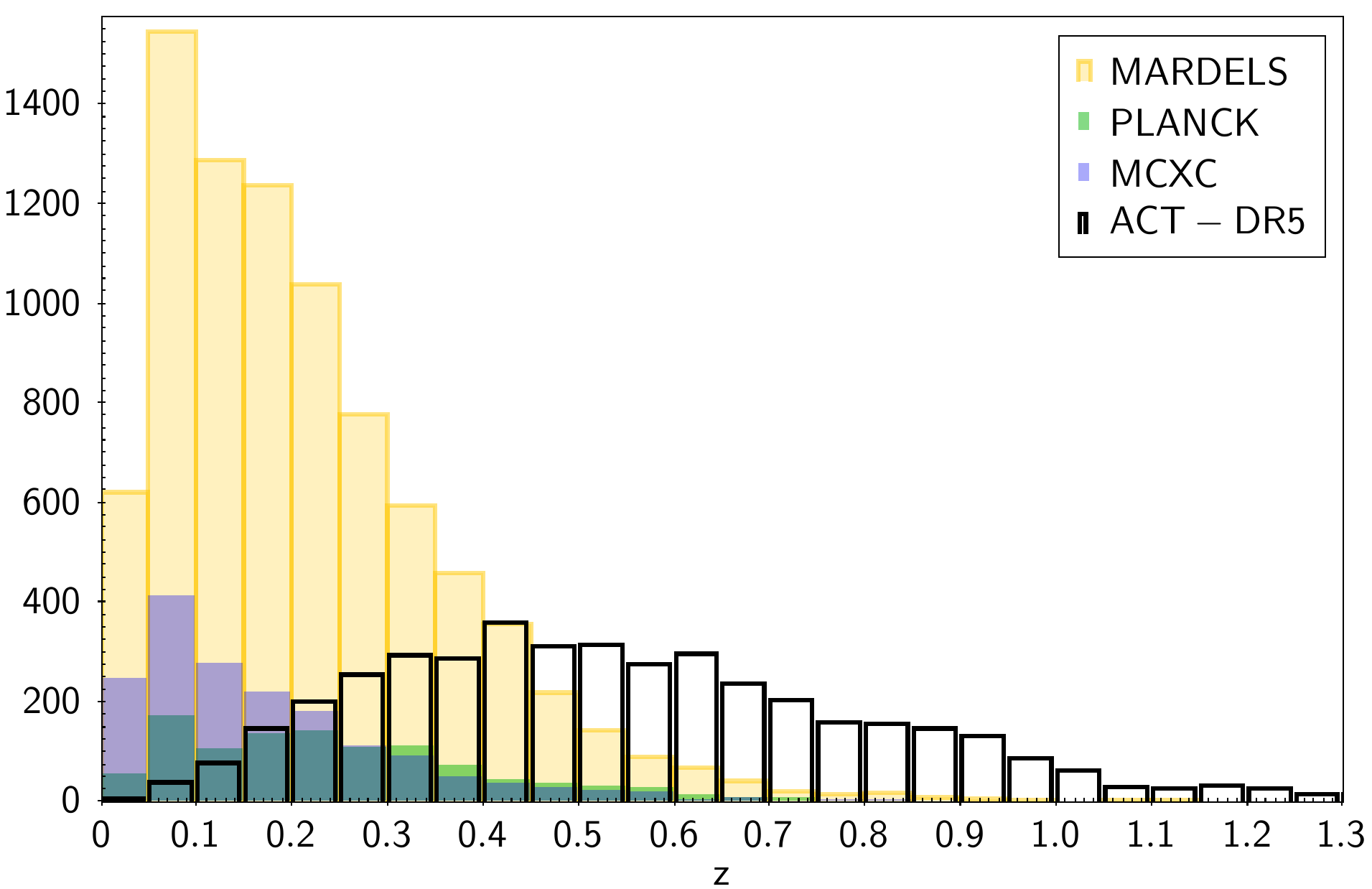}
\vskip-0.10in
\caption{Left: X-ray mass proxy $M_X$ versus redshift for the 90\% pure RASS-MCMF sample. Highlighted in blue are matches to the MCXC, Planck PSZ2 and the ACT-DR5 cluster catalogs. Right: Redshift distribution of the 90\% pure RASS-MCMF sample, as well as for MCXC, Planck PSZ2 and the ACT-DR5.
The RASS-MCMF catalog contains more clusters per redshift interval ($dN/dz$) than ACT-DR5 out to $z\sim0.4$ and more clusters overall than all three external cluster surveys put together.
}
\label{fig:redshiftandmass}
\end{figure*}

\subsection{Catalog definition}
\label{sec:CatalogDefinition}
We present here a clean cluster catalog of 90\% purity that is built from 2RXS by applying the MCMF algorithm to the Legacy Survey DR10 dataset.  The key contamination removal steps include 1) removal of sources with low significance optical counterparts (Section~\ref{sec:MCMFselection}), 2) removal of multiple detections of the same cluster (Section~\ref{sec:rejectingmultiples}) and 3) an additional point source rejection (Section~\ref{sec:additionalPSremoval}).  

Table~\ref{tab:catalogprop} contains the properties defining the 90\% pure sample as well as the 95\% and 99\% subsamples. From left to right in the table we present the sample purity and the number of clusters in the sample.  In addition, we present the \fcont\ selection threshold applied, the number of 2RXS candidates selected, the number of multiple detections rejected and the number of point sources rejected.

We consider point source rejection that impacts the sample selection at the subpercent level as small enough to be ignored in most studies. However we acknowledge the fact that some of these sources might indeed be of astrophysical interest like clusters with strong central AGN emission or star formation like the Phoenix cluster \citep{McDonald15}. In fact, the Phoenix cluster is likely the most famous source excluded by our point source rejection step, although it almost avoided the point source rejection threshold with $\log(10^{14}\lambda/M_X)=1.096$. Other known clusters with similar features such as A1835 and  Zwicky 3146 \citep{Allen92}, A2667 \citep{Rizza98} or CHIPS1356-3421 \citep{Somboon21} are retained in our sample. We therefore believe that the RASS-MCMF sample presented here represents an excellent resource for most galaxy cluster studies.

We emphasize that the 99\% pure sample has negligible contamination, and the point source rejection step plays a smaller role here than in the 90\% and 95\% pure samples.

\begin{table}
\centering
\caption{Properties of the three RASS-MCMF galaxy cluster samples. From the left are the sample purity, the final number of galaxy clusters, the \fcont\ selection threshold,  
the total number of 2RXS sources selected,  the number excluded due to
being multiple detections of the same source, and the number of sources excluded due to point source rejection.}
\label{tab:catalogprop}
\resizebox{\linewidth}{!}{%
\begin{tabular}{|c|c|c|c|c|c|}
\hline 
Sample & Number of & \fcont\ Selection & Number of & Rejected &  Rejected\\ 
Purity&  Clusters& Threshold  & Candidates & Multiples &  Point Sources\\ 
\hline 
90\% & 8465 & 0.17 & 11585 & 2070 & 1044 \\
95\% & 6924 & 0.11 & 9214  & 1838 & 446 \\
99\% & 5516 & 0.06 & 7352 & 1636 & 194 \\
\hline
\end{tabular} 
}
\end{table}

\subsection{Spectroscopic redshifts}
\label{sec:spectroscopicredshifts}
We estimate spectroscopic redshifts for the best optical counterpart identified using MCMF by employing public spectroscopic galaxy redshifts. The galaxy redshifts are drawn from a merged catalog of the SDSS~DR17 \citep{sdss17}, 2dFGRS \citep{2df}, 6dFGS \citep{jones09}, 2MRS \citep{2MRS} and the spectroscopic subset of GLADE+ \citep{GLADE}. As a first step, we match this catalog with the positions of successfully estimated BCGs using a maximum positional offset of 2~arcsec. 
As a second step, we search for all spectroscopic galaxies within 2 Mpc and $\vert z_\mathrm{cluster}-z_\mathrm{spec} \vert < 0.025(1+z_\mathrm{cluster})$.
From the selected galaxies, we derive the median redshift and finally derive the cluster redshift using all galaxies within $\left| \delta z\right|<0.015$ from the median redshift. In the case where BCG redshifts exist, we select galaxies within $\left|\delta\right| z<0.015$ from the BCG redshift. In our final cluster catalog, we only list spectroscopic redshifts based on at least two members or with a BCG redshift. 

In total we provide spectroscopic redshifts for $\sim53\%$ of the RASS-MCMF cluster sample, which reduces to 40\% when requiring five or more spectroscopic members. Requiring five or more spectroscopic members, we then find that the scatter between MCMF photo-z and spectroscopic redshift $(\delta z/(1+z))$ is well described by a Gaussian distribution with standard deviation of $\sigma=0.0048\pm0.0001$. Due to the various depths and redshift ranges covered by the different spectroscopic surveys we employ, the fraction of clusters with spectroscopic redshifts changes significantly over the footprint. In the area covered by SDSS, we are able to provide spectroscopic redshifts for 93.5\% of the RASS-MCMF clusters.

\subsection{Properties of the cluster catalog}
\label{sec:properties}
With 8,465(6,924 and 5,516) clusters in the RASS-MCMF 90\% (95\% and 99\%) purity sample, this catalog contains the largest ICM-selected cluster sample to date. By covering $\sim$25,000 deg$^2$ of extragalactic sky, the survey area covers $>$90\% of the sky that is not significantly impacted by high stellar density or high Galactic N$_\mathrm{H}$ column density. 

In the left panel of Figure~\ref{fig:redshiftandmass} we show the distribution of the 90\% pure RASS-MCMF sample in estimated mass $M_X$ (see equation~\ref{eq:xray_type_II}) and redshift. As an approximately all-sky survey, the RASS-MCMF sample has overlap with most previous cluster surveys. Restricting to the largest ICM-selected cluster catalog ACT-DR5 \citep{ACTDR5} and the two largest ICM-selected all-sky catalogs Planck PSZ2 \citep{PSZ2} and MCXC \citep{MCXC}, we find $\sim$2,000 clusters in common with RASS-MCMF; these are marked in blue in Figure~\ref{fig:redshiftandmass}. Each of the three surveys individually contains $\sim900$ clusters in common with RASS-MCMF. As visible by the mass range covered by the blue points with respect to the full sample, the RASS-MCMF sample reaches lower masses than the other surveys out to $z\sim$0.4. Above this redshift ACT-DR5 clusters probe to lower mass than RASS-MCMF, causing the overlap between the two samples to cover the full dynamic range probed by RASS-MCMF. 

In the right panel of Figure~\ref{fig:redshiftandmass} we show the redshift distribution of the RASS-MCMF sample with respect to the aforementioned samples. The RASS-MCMF redshift distribution peaks at $z\sim0.1$ and shows a strong decrease in the number of clusters with redshift as expected for an X-ray selected cluster sample. It also shows that RASS-MCMF significantly outnumbers each of the three samples. Thanks to the weak mass dependence of the SZE signature on redshift, ACT-DR5 exceeds RASS-MCMF in the number of clusters per redshift interval ($dN/dz$) above $z=0.45$ and can therefore be seen as complementary to the X-ray-based RASS-MCMF sample. 

In Figure~\ref{fig:RASS-MCMFz00501skydist} we show the sky distribution of RASS-MCMF clusters in the narrow redshift range $0.05<z<0.1$. This redshift slice contains $\sim$1,500 clusters, almost as many as the entire MCXC or Planck PSZ2 samples. At this redshift range the RASS-MCMF catalog allows one to nicely sample the large scale structure or so-called cosmic web, which in this figure is traced using the 2MASS Photometric Redshift Catalog galaxy density map \citep{2MPZ}. For highlighting the advances over previous RASS-based cluster surveys, we also show clusters from the REFLEX \citep{REFLEX} and NORAS \citep{NORAS} samples as magenta squares.

\begin{figure*}
\includegraphics[keepaspectratio=true,width=1.0\linewidth]{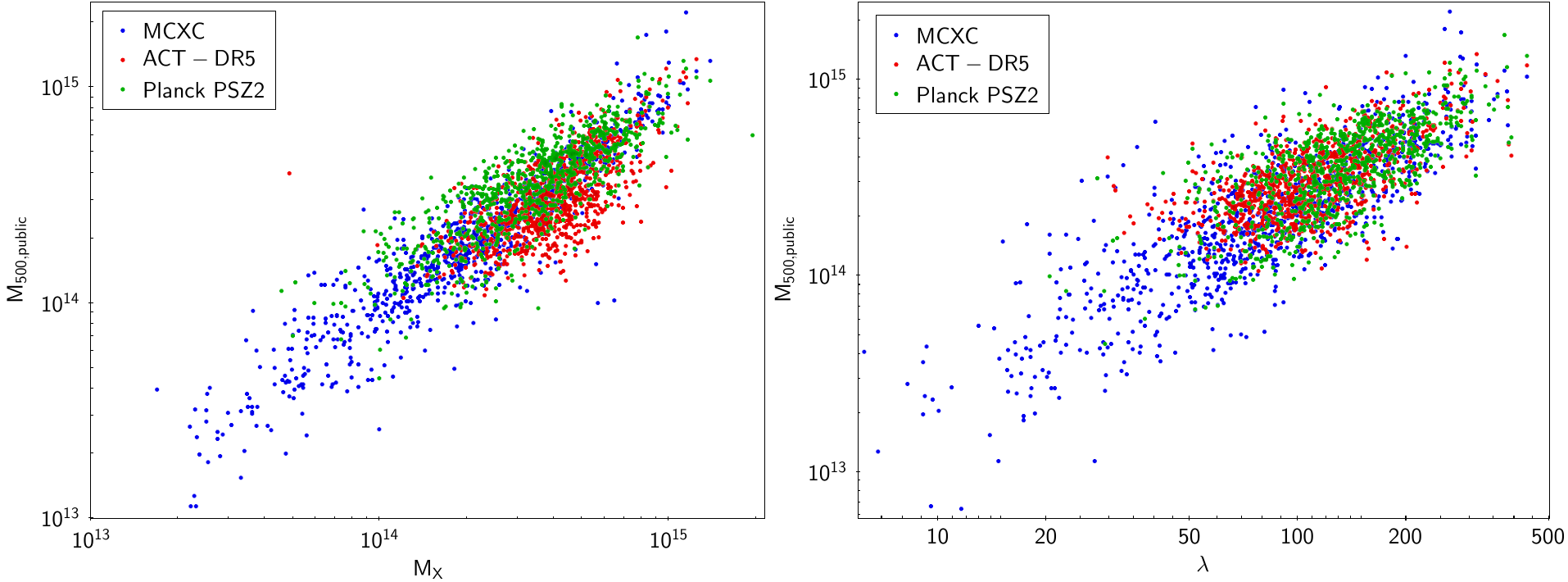}
\vskip-0.10in
\caption{Left: Comparison of the mass estimator $M_X$ employed here with mass estimates from ACT-DR5, MCXC and Planck for clusters in the 90\% pure RASS-MCMF sample. Right: Same but in comparison to richness.}
\label{fig:lambdamass}
\end{figure*}

\begin{figure*}
\includegraphics[keepaspectratio=true,width=1\linewidth]{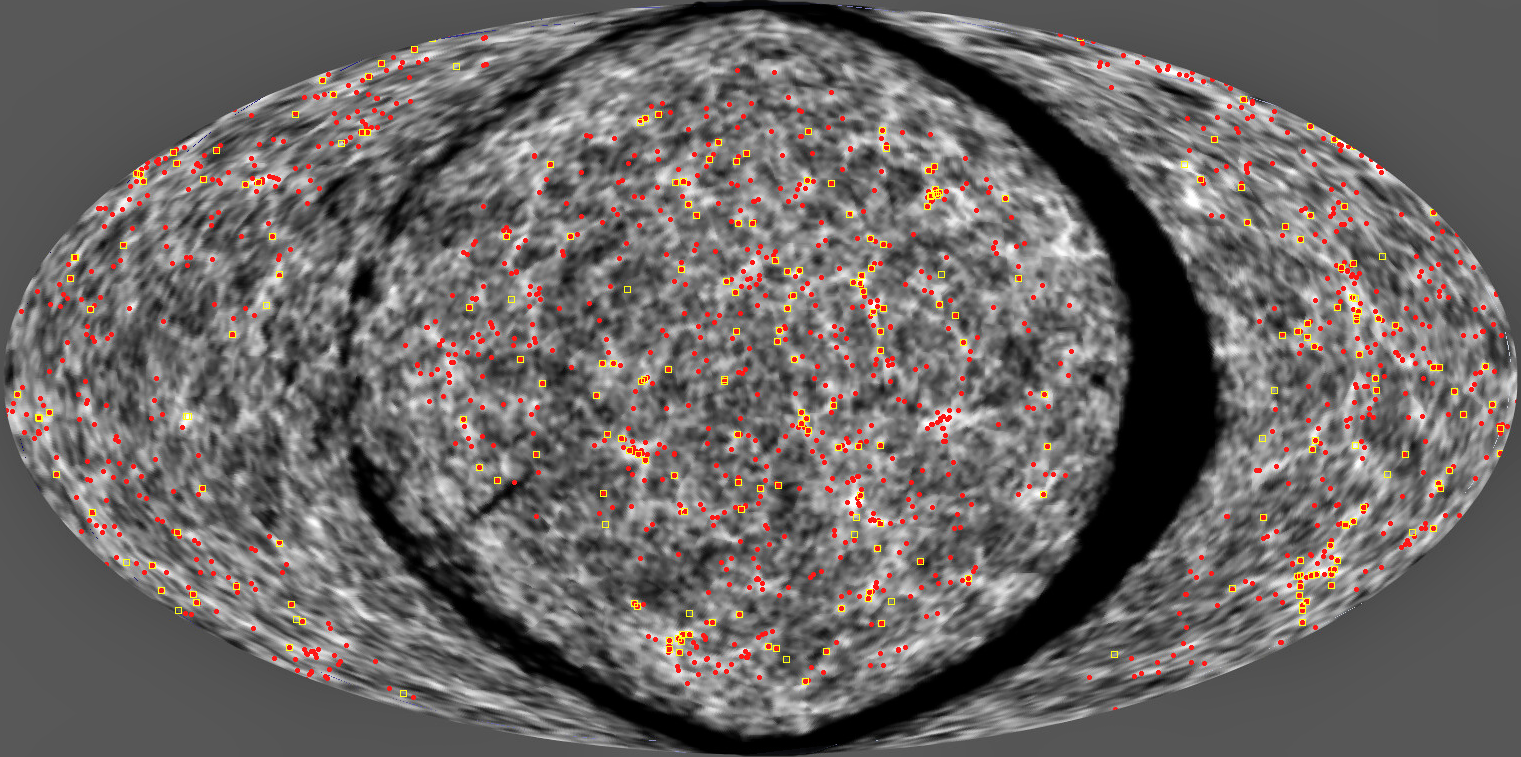}
\vskip-0.10in
\caption{Aitoff projection of the galaxy density at $0.07<z<0.1$ from the 2MASS Photometric Redshift Catalog \citep{2MPZ} centred on the North Galactic Pole. Red points show a $0.05<z<0.1$ redshift slice of the 90\% pure RASS-MCMF sample, containing $\sim 1,500$ clusters. For comparison, we show clusters over the same redshift range from the combined ROSAT-based NORAS and REFLEX catalogs (241 clusters) as yellow squares.}
\label{fig:RASS-MCMFz00501skydist}
\end{figure*}

From our previous work on MARD-Y3 \citep{Klein19} and subsequent work  on the validation of the selection function of that sample \citep{Grandis20}, we do expect reasonable scaling of the two mass observables ($M_X$, $\lambda$) of this sample with the underlying true halo mass. This is also illustrated in Figure~\ref{fig:lambdamass} which shows the scaling with masses taken from ACT-DR5, MCXC and Planck PSZ2 for the clusters in common. In the case of Planck clusters, we updated the masses using the correction found in \citet{Salvati22}. This plot highlights the large dynamic range in masses covered by RASS-MCMF, reaching well into the group mass regime at the low redshift end. We remind the reader that the low level of scatter for MCXC masses against $M_X$ shown in Figure~\ref{fig:lambdamass} is very likely due to the fact that both masses are derived from the same RASS data. When comparing the ACT-DR5 datapoints in both panels of Figure~\ref{fig:lambdamass}, one can see some indication of data points at $M_\mathrm{500,public}\approx 3\times10^{14} M_\odot$ scattering to either higher $M_X$ or lower  $M_\mathrm{500,public}$. Given that we do not see this in scaling with richness, this suggests that $M_X$ might be scattered high at higher redshifts. This might be evidence of Eddington and Malmquist bias coming into play at the low count rate regime. Alternatively the redfshift evolution assumed in the derivation of $M_X$ from count rate might be biased. 

This highlights the importance of a dedicated mass calibration of the RASS-MCMF sample including the modeling of the selection function. This has recently been done using HSC-SSP weak lensing in an MCMF-based X-Ray survey \citep{Chiu23}.  RASS-MCMF essentially covers all surveys with dedicated weak gravitational lensing programs, making such a weak lensing mass calibration a natural next step for this sample.

\section{Comparison to other cluster catalogs}
\label{sec:Validation}

In this section we investigate the RASS-MCMF catalog properties through comparison to external X-ray or SZE selected samples. These include: 
1) the eROSITA Final Equatorial-Depth Survey X-ray catalog (eFEDS),
%\citep[eFEDS][]{Brunner21}, 
2) the CODEX X-ray selected sample, 
%\citep{CODEX19}, 
3) the Planck-PSZ2 SZE selected catalog, 
%\citep{PSZ2},
4) the MCXC X-ray selected catalog,
%\citep{MCXC} 
5) the ACT-DR5 SZE selected catalog,
and
6) the MARD-Y3 X-ray selected catalog.
%\citep{ACTDR5}.

\subsection{eFEDS}\label{sec:eFEDS}
The extended ROentgen Survey with an Imaging Telescope
Array \citep[eROSITA;][]{Predehl21} observed during its performance verification phase a $\sim$140 deg$^2$ region with an average net exposure time of $\sim$1.3\,ks.  This survey was called the
 eROSITA Final Equatorial-Depth Survey \citep[eFEDS;][]{Brunner21}. eROSITA and its dedicated all-sky survey (eRASS) are the successors of ROSAT and RASS, showing much higher sensitivity and improved imaging resolution. eFEDS is therefore an ideal testing ground to investigate the expected purity and completeness of the RASS-MCMF sample. The eFEDS source catalog is divided into an unresolved source sample \citep{Brunner21} and an extended source sample \citep{Liu22a}, with corresponding optical confirmation presented in follow-on papers \citep{Salvato21,Klein22,bulbul22}. The majority of the RASS-MCMF clusters can be expected to be included in the sample of extended sources, however matches to clusters in the point source sample cannot be excluded because several hundreds of clusters in the eFEDS catalog do not fulfill the requirements to be included in the extend source sample \citep{bulbul22,Chiu22}.

Of the 90\% purity RASS-MCMF sample, only 39 lie within the eFEDS footprint.
Using a maximum separation of three arcminutes between 2RXS and eFEDS position we find 32 matches to the sample of eFEDS extended sources, 5 matches to eFEDS unresolved sources, and two sources do not have a match to any eEFDS source. 
Of the 5 matches to the unresolved sources, two are clearly galaxy clusters and were identified as such in the eFEDS papers. One detection is associated with the unrelaxed cluster eFEDS J085751.6+03103, where 2RXS finds two sources that can be associated with two clumps of galaxies, while the eFEDS detection pipeline classifies the system as one. Another RASS-MCMF source is dominated by a point source, that is likely a cluster member of eFEDS J084544.3-002914, which is located four arcminutes away from the 2RXS position. The remaining source associated with an unresolved eFEDS source and the two sources without a match to any eFEDS sources are likely random superpositions, which serve as contamination within the RASS-MCMF sample.  These 3 contaminating sources out of 39 total are in good agreement with the expected 10\% contamination of the RASS-MCMF sample.

\subsection{CODEX}
The COnstraining Dark Energy with X-ray cluster survey \citep[CODEX:][]{CODEX19} is based on the same ROSAT raw data as the 2RXS catalog in our work but uses a different source detection algorithm, namely a wavelet decomposition method \citep{Vikhlinin98}. A total of 24,788 X-ray sources were found over the $\sim$10,500~deg$^2$ of the SDSS BOSS footprint. To identify optical counterparts for clusters in the X-ray source list,
the CODEX team uses the redMaPPer optical cluster finder \citep{Rykoff14} run at the X-ray positions using SDSS photometric data. After obtaining redshifts and richnesses of possible optical counterparts, 
they perform a redshift dependent richness cut that is similar to that used in the MCMF study of 2RXS sources over the DES region \citep{Klein19}.  With this cut they produce what they estimate to be a 95\% pure cluster sample. Those X-ray sources making this cut are flagged as clean in the CODEX catalog. The publicly available CODEX catalog contains 10,382 sources of which 2,815 are flagged as clean. 
%We compare to this clean subsample in the discussion below.

Over the BOSS footprint we find $\sim$42,000 2RXS sources, 70\% more than the CODEX candidate list, which is a reflection of the different X-ray selection techniques used to construct the two candidate lists. 
Out of the list of CODEX confirmed sources, only 2,033 (72\%) have a match to 2RXS within a 3~arcminutes radius. From the matched sources we find $\sim85$\% of the sources to be part of the 90\% pure RASS-MCMF sample.

We test the purity of the clean CODEX subsample by repeating the test we performed for the 90\% purity RASS-MCMF sample on eFEDS (Section~\ref{sec:eFEDS}). We find 33 CODEX sources in the eFEDS footprint, of which 27 match to eFEDS extended sources, and one of the remaining 6 CODEX sources match to a cluster in the eFEDS unresolved source sample. 
The remaining CODEX sources typically match to a bright point source in the vicinity of a cluster, but showing offsets greater than 3.5~arcmin. 
The contamination noted for the clean CODEX sample would translate into an expectation of $\sim$1.6 false sources, which is small compared to the 5 we find with matches to point sources in eFEDS, suggesting that the purity of the clean CODEX sample is overestimated. The full CODEX sample over the eFEDS footprint contains 123 sources. Only 42 match to extended eFEDS sources and 6 to clusters in the point source catalog. This suggests that $\sim$60\% of the CODEX sources provided in the full CODEX sample are not X-ray selected clusters.

The comparison of total numbers and the redshift distribution over the CODEX footprint is also interesting. The 90\% purity RASS-MCMF sample contains 
$\sim$4,000 clusters over the CODEX footprint, while the 95\% pure RASS-MCMF sample contains $\sim$3,300. 
Thus, the 95\% purity RASS-MCMF sample provides 17\% more sources than CODEX, while the 90\% purity sample shows 50\% more sources than CODEX.
Part of this difference between CODEX and RASS-MCMF arises because the Legacy Survey data are deeper than SDSS, and therefore it is possible to confirm clusters at higher redshift in RASS-MCMF.

By cross matching CODEX clean sources with RASS-MCMF sources lying within 3~arcminutes, we find generally good agreement between the samples in the redshift range  $0.1<z_\mathrm{CODEX}<0.5$. We see the known bias at redshifts below $z_\mathrm{CODEX}=0.1$ \citep{clerc16} and some trend for redshift underestimation at high redshifts where SDSS imaging depth is reaching its limits for cluster member galaxy detection. We see only $\sim$1.9\% outliers in photometric redshifts for the crossmatched 90\% purity sample, and all but 2 outliers (0.1\%) show a redshift match between the CODEX redshift and the 2nd or third ranked RASS-MCMF counterpart redshift. This highlights the fact that outliers between both surveys are typically not caused by redshift measurement errors, per se, but by the choice of which optical structure to associate with an X-ray source. 

For sources with consistent redshifts we find a significant redshift trend in the ratio of richnesses, suggesting a factor two increase in $\lambda_\mathrm{CODEX}$ from redshift of z= 0.2 to z=0.55 relative to the RASS-MCMF richnesses. 
%This trend is highlighted in Figure~\ref{fig:CODEXrich} for the CODEX overlap with the 90\% purity RASS-MCMF sample. 
A similar redshift trend was reported in \cite{IderChithamCODEX2020}, where they remeasured richnesses for CODEX clusters in the SPIDERS subsample \citep{clerc16} using Legacy Survey DR8 data. Using a redshift dependent richness cut from \citet{Klein19} without accounting for this redshift trend would lead to an increasing contamination of the CODEX clean sample with redshift.

%\begin{figure}
%\includegraphics[keepaspectratio=true,width=1\linewidth]{MARDELS/2RXS_RASS-MCMFvsCODEXrichness.pdf}
%\caption{Ratio of CODEX richness ($\lambda_\mathrm{CODEX}$) to RASS-MCMF richness ($\lambda_\mathrm{RASS-MCMF}$) as a function of redshift for the CODEX crossmatched sample of the RASS-MCMF 90\% purity sample. The red line represents the running median ratio within a redshift window of $\Delta z=0.05$.}
%\label{fig:CODEXrich}
%\end{figure}

\subsection{Planck-PSZ2}

The Planck PSZ2 \citep{PSZ2} catalog is the second catalog of SZE selected sources derived from the Planck dataset.
It contains 1,653 sources with an expected purity of at least 83\%; of these, 1,203 are considered to be optically confirmed. Only 1,261 PSZ2 sources fall within our RASS-MCMF footprint. The median positional uncertainties of the PSZ2 sources is comparably large ($\approx$2.4~arcmin); we therefore adopt a cross-matching radius of 5 arcminutes.
We find 881 matches to 2RXS sources, of which 842 have a match to the 90\% purity sample and an additional 12 systems make the \fcont\ selection threshold but are rejected as likely AGN or stellar sources. From the remaining 27 PSZ2 sources matched to 2RXS, only 10 systems have a measured redshift and can be considered to be optically confirmed. We visually inspect those systems and find two obvious missed clusters (PSZ2 G047.71-59.47, PSZ2 G046.26-70.47), both located in the patchy $gi$-band part of the footprint.
The remaining 8 systems are either at large separation from the 2RXS source ($>300$ arcsec) or are potentially low richness systems or perhaps chance superpositions. 
From repeating the matching with the random catalogs, we estimate a $\sim15$ chance of matching physically unassociated PSZ2 and 2RXS systems. This provides an explanation for the remaining 8 systems not being confirmed.

From the 854 sources that either have a match with the 90\% purity sample or make the \fcont\ cut but are then excluded as likely AGN, 761 have a PSZ2 redshift. Only 8 of these sources show redshift offsets of $\Delta z/(1+z)>0.1$, which are very large by galaxy cluster standards. In all these cases, the RASS-MCMF redshift seem to be the correct redshift to the 2RXS candidate. In three cases the matched PSZ2 cluster corresponds to another cluster and refers to the 2RXS candidate in question. Finally we find five cases where the redshifts listed in PSZ2 likely need to be reconsidered. 
One noteworthy case of a likely wrong redshift is PSZ2~G181.71-68.65 (ACO~305) with $z_\mathrm{PSZ2}=0.1529$, which is likely at $z_\mathrm{MCMF}=0.293$. We list the five cases with discrepant redshifts in Appendix~\ref{app:PSZphotozoff}.

We summarise the comparison as follows: we find 842 matches to the 90\% purity sample, with only $\sim5$ being redshift outliers where Planck likely lists incorrect redshifts. We find two cases where the local quality of the optical data do not allow us to confirm the clusters.

\subsection{MCXC}
The Meta-Catalog of X-Ray Detected Clusters of Galaxies \citep[MCXC][]{MCXC} is a collection of various ROSAT-based cluster surveys, including RASS-based (BCS, CIZA, MACS, NEP, NORAS, REFLEX and SGP) as well as samples obtained from serendipitous detections within pointed ROSAT PSPC observations. Within a distance of 2.5~arcmin we find 985 matches to the 90\% pure RASS-MCMF sample and an additional 45 sources making the \fcont\ cut but marked as possible point sources or multiple detections. 

The majority of these systems ($\sim90$\%) are matches to RASS based surveys. Further, we find 79 matches to 2RXS that do not make the RASS-MCMF selection thresholds. Of those, 27 (33\%) are matched to the NEP survey \citep{Henry_2006}. Visual inspection of these suggest that the majority ($\sim67\%$) of these matches are not clusters, while two have an \fcont\ close to the selection threshold. A list of NEP sources likely misclassified as clusters is provided in Appendix~\ref{app:NEPmisc}. This further stands in strong contrast the matches to the SGP survey \citep{Cruddace_2002}, for which we confirm 48 out of 49 clusters or the BCS sample \citep{Ebeling1998} for which confirm all matches. From the remaining matches, 41 are RASS based with 37 of those coming from the REFLEX or NORAS surveys, which share flux limit and source identification methods. Visual inspection of these sources using optical and auxiliary X-ray data indicate that $\sim 1/3$ of those matches are either point sources or strongly dominated by point sources. Those sources typically lie at the high luminosity and redshift regime of the non-confirmed systems, where the optical cleaning in our catalog should have a smaller impact on the completeness. 
Where we do expect our selection methods to lead to the loss of real systems is predominantly at low redshifts and masses. Here the LS photometry and red sequence techniques hit their limit with respect to galaxy size and contrast to non-collapsed structures.
This is reflected in the fact that 30\% of the missed systems are groups at $z<0.02$. The median of all unconfirmed systems is $z=0.055$. In total we find four clusters, where visual inspection, cluster redshift and X-ray luminosity would lead us to expect them to be detected in our sample. Two of them fall into the patchy region where only $gi$-band data are available. A third case is RXC~J0105.0+0201 (z=0.197), which lies projected behind the nearby (z=0.006) galaxy IC1613. It is likely in this case that the $>15$ arcmin size of the galaxy resulted in over-subtraction of the background in the Legacy Survey DR10 images and may also have impacted photometric calibration at the location. 

Comparing redshifts between MCXC and our 90\% pure sample, we find that redshifts agree for 98\% of the cases within $\Delta z<0.05$. Investigating the remaining 21 sources with larger offsets, we find 16 cases where RASS-MCMF redshifts are correct as confirmed by publicly available spectroscopic redshifts. The remaining five show at least two clusters along the line of sight, all indicated with a second strong counterpart with \fcont$<0.1$. In all except one case the primary counterpart listed in RASS-MCMF is the better or an equally good counterpart. In only one case, RXC~J1036.6-2731, at z=0.013, the second ranked counterpart from RASS-MCMF seems to be the preferred counterpart. 

We summarise the comparison to MCXC as follows, we find $\sim$1000 clusters in common with the 90\% purity RASS-MCMF sample. We further find 79 positional matches which do not make our selection cuts. From those, the majority of sources are either point sources, point source dominated clusters or low redshift and low mass systems. The small number of missed massive clusters fall either in the $gi$-band footprint or are affected by foreground galaxies.
For redshifts we find generally good agreement between the catalogs, the majority of the 2\% redshift outliers are caused by MCXC listing a foreground source rather than the true cluster at higher redshift.

\subsection{ACT-DR5}
The ACT-DR5 cluster catalog \citep{ACTDR5} is currently the largest SZE-selected cluster catalog, containing 4,195 clusters over a sky area of 13,211~deg$^2$. More than 98\% of the sources are within the Legacy Survey DR10 footprint. Using a matching radius of 2.5~arcmin, we find 1,074 matches to 2RXS sources, of which 915 appear in the 90\% pure RASS-MCMF sample, 23 were excluded as likely point sources, %\fcont$<0.17$ 
and 136 simply did not make the \fcont\ selection threshold meant to exclude likely random superpositions.  Many of the SZE selected clusters are at $z>0.5$, where ACT probes significantly lower masses than our catalog.

Among the cross-matched cluters, we find good redshift agreement for 98.5\% of the cases. From the remaining sources 80\% have a match to the second ranked source in our sample, suggesting multiple clusters along the line of sight as the main source for redshift discrepancies. In contrast to our previous exercise on MCXC, ACT-DR5 redshifts tend to lie above our redshifts.  This is likely a consequence of the different redshift dependencies of the cluster selection methods. 
%In the case of ACT this function is rather flat with redshift while it is strong for X-ray selected samples. 
Because of the approximately redshift independent mass threshold in its survey, the ACT-DR5 team prefers the richest system along the line of sight at the position of their SZE candidate.  Our RASS-MCMF selection depends on richness and redshift, identifying the highest significance optical counterpart along the line of sight toward the X-ray selected candidate.  In principle, it would even be possible for the SZE and X-ray selected systems to be different along a line of sight, in which case the redshifts could be correctly assigned and still not agree.

Of greater scientific interest than the small fraction of redshift outliers is the question what 2RXS selected clusters matched to ACT-DR5 do we miss in our 90\% pure RASS-MCMF sample. %In Figure~\ref{fig:act} we show the distribution of 2RXS matches to ACT-DR5 clusters in mass versus redshift taken from ACT-DR5 measurements. The matches that do not make the selection cut to make it into the 90\% pure RASS-MCMF sample are highlighted in blue. Given the sample size and source density, about half (73 sources) of the positional matches to ACT can be expected to be chance matches. 
%Highlighted with red circles are three ACT clusters with masses and redshifts expected to make the selection threshold.
Looking at the ACT masses and redshifts of those systems unconfirmed systems, we find three ACT clusters that are massive enough that they should be well above our selection thresholds. 
Of those three, two-- ACT-CL J0105.0+0201 and ACT-CL J2248.5-1606-- were already found to be missing in comparison to the MCXC catalog. In case of ACT-CL J2248.5-1606 being one of two two clusters lying in the patchy region where only $gi$-band data are available. The last cluster-- ACT-CL J1355.1+0430 ($z=0.185$)-- is a more complicate case. Our analysis finds an additional cluster at $z=0.81$ with richness $\lambda=68$ at the ACT position, suggesting this is the correct counterpart compared to the $\lambda=17$ system at lower redshift. In addition, we find a good QSO point like counterpart for this 2RXS source. Placing the ACT cluster at $z=0.81$ would likely put the cluster out of reach for 2RXS. The ACT catalog also contains a note about this cluster, indicating this it might be a projected system. The good AGN counterpart further supports that the 2RXS match is indeed a point source instead of a massive cluster. The remaining ACT matches show typically low masses, where scatter in the mass-richness relation could explain why these systems do not make our selection cut.

\subsection{MARD-Y3}
The MARD-Y3 catalog \citep{Klein19} is the result of the systematic MCMF follow-up of 2RXS over the DES footprint using the DES-Y3 data set \citep{DESDR1}. It therefore shares similarities in method, the same X-ray catalog and in part imaging data. Key differences are additional (deeper) imaging data in case of LS~DR10 and improved calibration at low redshifts. Furthermore, the richness distribution for the contaminants are drawn over different areas (DES or full Legacy Survey DR10). The method to merge multiple detections also differs for the two catalogs.

To match the catalogs, we directly match the original 2RXS sources. This avoids ambiguity due to different choices in the rejection of multiple detections of the same cluster. 
For an \fcont\ selection threshold of 0.17, we find 2,626 (RASS-MCMF) and 2,599 (MARD-Y3) 2RXS sources over the same footprint, but only 81\% of them make the selection cut in both surveys. That is, there are $\sim$500 sources in each catalog that don't appear in the other catalog. The reasons for this are 1) different richness thresholds at a given redshift for the same \fcont=0.17 selection and 2) scatter between the cluster richnesses extracted from DES and from LS~DR10.

The richness thresholds are different because the
average RASS exposure time over the DES footprint is higher than that over the full RASS-MCMF footprint. This leads to an increase in the probability that an optical cluster of a given richness is also a X-ray selected cluster to be higher over DES than on average over the full RASS-MCMF area. Consequently the minimum richness needed at fixed \fcont\ is lower for MARD-Y3 than for RASS-MCMF. In addition, the large solid angle and the improved coverage over SDSS in the RASS-MCMF sample allow for a better calibration of the cluster models as well as better statistics to model the \fcont\ selection toward low redshifts. 
%The crossover in redshift dependent richness selection is at $z\sim0.1$.
As a result, RASS-MCMF systems not in MARD-Y3 but with \fcont$<0.17$ are predominantly at $z<0.2$, while MARD-Y3 sources not in RASS-MCMF exhibit a broad redshift range. 

A greater concern would be if there were clusters that have a high enough richness to not be affected by the \fcont\ threshold but nevertheless appear in only one catalog.
%show a richness high enough that they have to lie above the selection threshold of the other catalog.
Looking at the richness scatter and the richness thresholds given \fcont\ we find two MARD-Y3 clusters and 12 RASS-MCMF clusters that fit into this category. We find that both MARD-Y3 clusters lie close to the nearby galaxies NGC~300 (z=0.0005) and IC~1683 and therefore likely suffer from masking in the Legacy Survey.
Similarly ten out of the twelve sources in RASS-MCMF show a lack of data in DES-DR3 at the cluster position. One of the remaining clusters is MACSJ0257.6-2209 \citep{Ebeling_2001}, that was discussed as a special case in the MARD-Y3 catalog paper \citep{Klein19} as a rare case where the local DES photometry was impacted by an error in the photometric zeropoint estimation.
For sources making the \fcont\ threshold in both samples we find 98.5\% of the cases have consistent redshifts, and that the remaining 1.5\% do have a match with a lower ranked optical counterpart in the other survey.

We summarise the comparison with MARD-Y3 as follows: for the same underlying 2RXS source and the same \fcont\ threshold we find 81\% overlap between both samples. Sources appearing in just one of the samples can be explained by the different selections in richness as function of redshift and by scatter between the two richness measurements.

%\begin{figure}
%\includegraphics[keepaspectratio=true,width=1\linewidth]{MARDELS/MARDELSxACT_massz_v3c.pdf}
%\caption{Mass versus redshift distribution of ACT-DR5 clusters matched to the full 2RXS catalog within the optical footprint. Clusters making the selection cuts of the 90\% pure RASS-MCMF sample are highlighted in gray, unconfirmed systems are shown in black. Unexpectedly unconfirmed systems are marked with red circles while sources falling below the gray line are expect to be highly likely chance super positions.}
%\label{fig:act}
%\end{figure}

%%%%%%%%%%%%%%%%%%%%%%%%%%%%%%%%%%%%%%%%%%%%%%%%%%
%%% COSMO forecast 
%%%%%%%%%%%%%%%%%%%%%%%%%%%%%%%%%%%%%%%%%%%%%%%%%%
\section{Cosmological forecast for RASS-MCMF$\times$DES}
\label{sec:Forecast}

Large, well understood cluster samples have been pursued over the last two decades largely because of the information they contain about the underlying physical processes responsible for the cosmic acceleration \citep{Haiman01}.  The RASS-MCMF sample together with the recent developments toward an accurate understanding of the HMF over a broad range of cosmologies \citep[e.g.,][]{Bocquet20} create a situation where all requirements for precise and accurate cluster abundance cosmological studies are met \citep[see discussion in][]{Mohr05}.  

To test the potential constraining power of the RASS-MCMF cluster sample when combined with DES weak lensing, we apply a cluster cosmology analysis code to representative mock datasets.  The cosmology analysis code has been developed for the study of South Pole Telescope SZE selected clusters \citep[][in prep]{Bocquet15,Bocquet19} in combination with gravitational weak lensing data.  It is written as a \textsc{CosmoSIS} module \citep{cosmosis}.

To enable this RASS-MCMFxDES forecast, we have extended this code to work on X-ray selected cluster samples extracted from all-sky X-ray surveys like those from ROSAT and eROSITA. In the following subsections we review the analysis method and the mock observations and then present the parameter constraint forecasts. 

\subsection{Cluster cosmology analysis method}
\label{sec:cosmointro}

The likelihood is estimated as a multi-variate Poisson likelihood using the expected number of clusters given the observables. The observables to consider for each cluster are the X-ray count rate $\hat\eta$, the richness $\hat\lambda$ and the redshift $z$.  Thus, our likelihood is closely related to that of \cite{Chiu23}:

\begin{equation}\label{eq:lnlike}
\begin{split}
    \ln \mathcal{L(\mathbf{p})} = & \left[ \sum_i \ln \left( C(\hat\eta, z) \dfrac{\mathrm{d}N(\hat\eta, \hat\lambda, z| \mathbf{p})}{\mathrm{d}\hat\eta\mathrm{d}\hat\lambda\mathrm{d}z} \right)  \bigg|_{i\mathrm{-th cluster}}\right.  \\
     & \left. - \int\displaylimits_{z_{\rm min}}^{z_{\rm max}}\mathrm{d}z \int\displaylimits_{\hat\eta_{\rm min}}^{\hat\eta_{\rm max}}\mathrm{d}\hat\eta \int\displaylimits_{\hat\lambda_{\rm min}(z)}^{\infty} \mathrm{d}\hat\lambda\, C_\mathrm{HOF}(\hat\eta, z) \dfrac{\mathrm{d}N(\hat\eta, \hat\lambda, z| \mathbf{p})}{\mathrm{d}\hat\eta\mathrm{d}\hat\lambda\mathrm{d}z} \right],
\end{split}
\end{equation}

\noindent
where $C(\hat\eta, z)$ and $C_\mathrm{HOF}(\hat\eta, z)$ represent the completeness function for each cluster $i$ and for the ensemble, respectively (see Section~\ref{sec:completeness}), $\hat\lambda_{\rm min}(z)$ is the minimum observed richness of a cluster at redshift $z$ (derived from the \fcont\ threshold) and $\frac{\mathrm{d}N(\hat\eta, \hat\lambda, z| \mathbf{p})}{\mathrm{d}\hat\eta\mathrm{d}\hat\lambda\mathrm{d}z}$ is the halo-observable function or HOF. 

We note two differences with respect to the likelihood from \cite{Chiu23}: (1) our selection function includes a maximum count rate, $\hat\eta_\mathrm{max}$, and (2) we use only the likelihood of the number counts of our mock RASS-MCMF clusters, excluding the ``mass calibration'' likelihood (right-most term in their Eq.~6). 
To include mass information from the DES shear \citep{GattiDESWL21} and photo-z \citep{MylesDESPHZ21} based weak lensing mass calibration, we adopt priors on the observable mass scaling relation parameters that come from posteriors derived from a separate, ongoing MARD-Y3$\times$DES analysis (Singh et al. in prep.). The redshift range we assume for the analysis is $z_{\rm min}=0.01$ and $z_{\rm max}=1.1$, with the maximum count rate fixed at $\hat\eta_{\rm max}=13$.

Similarly to \cite{Chiu23}, the HOF is calculated from the halo-mass function \citep[HMF; ][]{Tinker08}, 
%accounting for the comoving volume at different $z$. 
using the observable mass relations for the intrinsic richness $\lambda$ and count rate $\eta$.  Appropriate convolutions are carried out to model the intrinsic and sampling or measurement scatter of the two observables. 

\subsubsection{Observable mass relations}
\label{sec:observablemass}

The underlying richness observable mass relation $\lambda-M-z$ has the form
\begin{equation}\label{eq:lambdamass}
    \left< \ln(\lambda | M,z)\right> = \ln A_\lambda  + B_\lambda \ln\left(\frac{M}{M_\mathrm{piv}}\right) + C_\lambda \ln\left(\frac{1+z}{1+z_\mathrm{piv}}\right),
\end{equation}
\noindent
where $M_\mathrm{piv}=1.6\times10^{14}h^{-1}$M$_\odot$ and $z_\mathrm{piv}=0.25$ are chosen to reflect the median mass and redshift of our mock cluster catalogue.  We adopt a log-normal intrinsic scatter in $\lambda$ at fixed mass and redshift that is the same for all redshifts and masses of
\begin{equation}
    \sigma_{\ln\lambda} = (\mathrm{Var}[\ln\lambda|M,z])^{\frac{1}{2}}. 
\end{equation}
Furthermore, we model the sampling noise on the expectation value of the richness $\lambda$ for a given mass and redshift as a Poisson distribution in the Gaussian limit. That is,
\begin{equation}
    P(\hat{\lambda}|\lambda) = \frac{1}{\sqrt{2}\delta_\lambda} \exp\left(-\frac{(\hat{\lambda} - \lambda)^2}{2\delta_\lambda^2}\right),
\end{equation}
\noindent
where $P(\hat{\lambda}|\lambda)$ is the distribution of the observed richness $\hat{\lambda}$ given the measurement uncertainty $\delta_\lambda = \lambda^{1/2}$.

The underlying count rate observable mass relation $\eta-M_{500}-z$ has the form \citep{Grandis19,Chiu22}
\begin{equation}\label{eq:etamass}
\begin{split}
    \left<\ln \left( \frac{\eta}{\mathrm{counts/sec}} \Bigg| M, z\right) \right> =& \ln A_\eta 
    - 2\ln\left( \frac{D_\mathrm{L}(z)}{D_\mathrm{L}(z_\mathrm{piv})} \right)\\
    +&\left[ B_\eta + \delta_\eta \ln \left( \frac{1+z}{1+z_\mathrm{piv}} \right)  \right] \ln\left( \frac{M}{M_\mathrm{piv}} \right) \\  +&\gamma_\eta  \ln\left( \frac{1+z}{1+z_\mathrm{piv}} \right) +2\ln \left( \frac{E(z)}{E(z_\mathrm{piv})} \right),
    %+ \ln (c_\mathrm{eRR}), 
    %+ \ln(b_\eta (M, z)),
\end{split}
\end{equation}
\noindent
where $D_\mathrm{L}$ is the luminosity distance, $z_\mathrm{piv}$ and $M_\mathrm{piv}$ are the pivot redshift and mass (same as for the $\lambda-M-z$ scaling relation), and $E(z)=H(z)/H_0$ is the expansion history of the Universe. 
The form of this observable mass relation allows for the mass trend to evolve with redshift.  It deviates from that presented in \cite{Chiu23} through the missing eROSITA based bias factor $b$ that we discuss further in Section~\ref{sec:mocks}.

As with the richness, we adopt 
a log-normal intrinsic scatter that is the same at all
redshifts and masses 
\begin{equation}
    \sigma_{\ln\eta} = (\mathrm{Var}[\ln\eta|M,z])^{\frac{1}{2}}. 
\end{equation}
Similarly, we model the $\eta$ measurement uncertainty as a Poisson probability with a mean corresponding to the expected number of photons $n_\gamma=\eta*t_\mathrm{exp}$, which corresponds to the expected count rate times the exposure time.  Explicitly,
\begin{equation}
    P(\hat{\eta}|\eta,t_\mathrm{exp}) = \frac{1}{t_\mathrm{exp}}
    \frac{n_\gamma^{\hat n_\gamma} e^{-n_\gamma}}
    {\hat n_\gamma !},
\end{equation}
\noindent
where $n_\gamma=\eta t_\mathrm{exp}$ and 
$\hat n_\gamma=\hat\eta t_\mathrm{exp}$.

Because the third observable, the cluster redshift $z$ has a high accuracy and precision ($\sigma(\delta z/(1+z)\sim0.005$; see Figure~\ref{fig:redshifts} and associated discussion in Section~\ref{sec:spectroscopicredshifts}) we do not model the redshift measurement uncertainty.
%\noindent

 As can be seen in Fig.~\ref{fig:footprint}, the exposure time varies across the sky from 100~s to over 3,000~s, and thus the transformation from $\eta$ to counts or photons also varies and impacts the Poisson measurement noise, which is needed to calculate the HOF. To account for this, we build a sequence of HOFs for different values of the exposure time, accounting for the survey solid angle at an exposure time by using the exposure time distribution of the RASS$\times$LS-DR10 sky. Given an exposure time, particular values of the HOF $\frac{\mathrm{d}N(\hat\eta, \hat\lambda, z| \mathbf{p})}{\mathrm{d}\hat\eta\mathrm{d}\hat\lambda\mathrm{d}z}$ are extracted from the sequence using interpolation.  
 
 Explicitly, for each cluster the exposure time $t_\mathrm{exp}$ at the cluster sky location is employed in extracting the appropriate value of the HOF.  The last term in the cluster counts likelihood (equation~\ref{eq:lnlike}) is evaluated using the sum of the sequence of exposure-time dependent HOFs.

\subsubsection{X-ray completeness function}
\label{sec:completeness}

The X-ray completeness function appears in our cluster abundance likelihood, because the X-ray selection on the 2RXS catalog is made using the existence likelihood EXI\_ML, while the primary observable mass relation is modeled using the count rate, which is directly related to the cluster X-ray flux and, given the redshift and cosmological parameters, also the cluster X-ray luminosity.  The existence likelihood of a cluster depends strongly on its flux or count rate, but it also has important dependencies on the ICM distribution (angular size and morphology of X-ray emission) as well as survey parameters such as the background and exposure time.  

Conveniently, the 2RXS sample has measured count rates $\hat\eta$ and existence likelihoods EXI\_ML for all objects, and the sample is drawn from the full range of exposure times and associated backgrounds. 
Previous X-ray cosmological analyses have used measured observables from the catalog to empirically defined the relationship between the count rate and the existence likelihood \citep{Klein19} or extent likelihood \citep{Chiu23} with good success, and therefore we proceed along this route for this forecast.

%To model the completeness function, we first need to account for the 2RXS existence likelihood threshold at EXI\_ML~$\geq6.5$. For this, 
We model the $\hat\eta$-EXI\_ML scaling relation needed for the completeness model therefore as follows
\begin{equation}\label{eq:etaeximl}
\begin{split}
    \left<\ln(\hat\eta|
    \mathrm{EXI\_ML}, t_\mathrm{exp}, \mathrm{bkg})\right> 
    =& \ln A_\mathrm{C}+ B_\mathrm{C}\ln\mathrm{EXI\_ML} \\
&+ C_\mathrm{C} \ln t_\mathrm{exp}+ D_\mathrm{C}\ln\mathrm{Bkg},
\end{split}
\end{equation}
\noindent
where $t_\mathrm{exp}$ is the exposure time and $\mathrm{Bkg}$ is the count rate of the background $\mathrm{Bkg} = \mathrm{Background}/t_\mathrm{exp}$.  This relation is modeled using a log-normal intrinsic scatter in $\hat\eta$ of
\begin{equation}
    s_{\ln\hat\eta} = (\mathrm{Var}[\ln\hat\eta| \mathrm{EXI\_ML}, t_\mathrm{exp}, \mathrm{Bkg} ]^\frac{1}{2})
\end{equation}
\noindent
%Daniel:  changes from log to ln as we discussed.

We extract the best-fit values using the RASS-MCMF clusters with a value of EXI\_ML close to the threshold value of 6.5. We find 
\begin{equation}
\begin{split}
%%% Switching from log_10 to ln
A_\mathrm{C} =\ & 1.802 \pm 0.061\\ %%%%  this is A
B_\mathrm{C} =\ & 0.328 \pm 0.002\\
C_\mathrm{C} =\ & -0.271 \pm 0.002 \\
D_\mathrm{C} =\ & 0.168 \pm 0.004\\
s_{\ln\hat\eta} =\ & 0.1819 \pm 0.0021.
\end{split}
\end{equation}

With this relationship between $\hat\eta$ and EXI\_ML, a selection threshold in the observed existence likelihood $\ln\mathrm{EXI\_ML}$ introduces a selection  in the observed count rate $\ln\hat\eta$ that is an error function. Therefore, we model the completeness function $C(\hat\eta, z)$ as
\begin{equation}\label{eq:complfunc}
    C(\hat\eta, z) = \frac{1}{2} \left( 1+ \mathrm{erf}\left( \frac{\ln\hat\eta - (\ln\hat\eta_{50} + \delta\ln\hat\eta_{50})}{\sqrt{2}s_{\ln\hat\eta}} \right)\right),
\end{equation}
\noindent
where erf() is the error function with the scaling factor $s_{\ln\hat\eta}$, where $\hat\eta_{50}$ is the count rate which has 50\% completeness ($\mathrm{EXI\_ML}=6.5$), 
and $\delta\ln\hat\eta_{50}$ is a parameter that allows one to model deviations from this expected threshold count rate during the cosmological analysis. Similarly, during the cosmological analysis we fit for the parameter $s_{\ln\hat\eta}$ scaling parameter. Depending on the priors adopted on these two parameters during the cosmological analysis, it is possible to self-calibrate the completeness function \citep[see discussion in][]{Chiu23}.

As already noted, Equation~(\ref{eq:lnlike}) contains two different forms of the completeness functions: $C(\hat\eta, z)$ and $C_\mathrm{HOF}(\hat\eta,z)$. The former is the completeness function appropriate for a particular cluster with a given observed count rate, exposure time and background.  The latter $C_\mathrm{HOF}(\eta,z)$ is the completeness of the full HOF used in the last term of the likelihood (see equation~\ref{eq:lnlike}).  This function is constructed first for each member of the sequence of HOFs created for different exposure time ranges (each corresponding to an equal width in $\Delta\log(t_\mathrm{exp})$).  We adopt the mean exposure time $t_\mathrm{exp}$ for each member of this sequence, because the exposure time width of each bin is sufficiently small.  To include the background Bkg dependence, we estimate a Bkg-weighted average completeness function $C_{\mathrm{HOF}_i}(\hat\eta,z)$ for exposure time bin $i$
\begin{equation}
    C_{\mathrm{HOF}_i}(\hat\eta, z) = \dfrac{\sum\limits_j p_j C_{\mathrm{HOF}_{i,j}}(\hat\eta, z)}{\sum\limits_{j} p_j}
\end{equation}
\noindent
where $C_{\mathrm{HOF}_{i,j}}$ is the HOF of the $i$-th $t_\mathrm{exp}$ bin and the $j$-th Bkg bin.  The factor $p_j$ is the weight of the $j$-th Bkg bin.

\begin{table}
    \caption{Summary of forecast parameters. First column corresponds to the name of the parameter, second column to the input value for the creation of the mock and third column corresponds to the priors adopted for the cluster cosmology analysis. The last two columns show the posteriors and uncertainties ($1\sigma$) for the two cosmologies: $\lambda$CDM and $w$CDM.}
    \label{tab:mocksNcosmology}
    \centering
    \resizebox{\linewidth}{!}{%
    \begin{tabular}{c c c c c}
        \hline
        %\hline
         Param. & Mock & Prior & \multicolumn{2}{c}{Posterior}\\
         & Input & & $\Lambda$CDM & $w$CDM \\
         \hline
         \multicolumn{3}{c}{The $\lambda-M_{500}-z$ scaling relation (Eq.~\ref{eq:lambdamass})}\\
         \hline
         $A_\lambda$ & 55.5 & $\mathcal{N}(55.5, 2.235^2)$ &  $55.9\pm 1.7               $& $54.7\pm 1.6                 $\\
         $B_{\lambda}    $ & 1.0 & $\mathcal{N}(1, 0.101^2)$ &  $1.044^{+0.049}_{-0.057}            $ & $1.012^{+0.047}_{-0.054}           $\\
         $C_{\lambda}    $ & 0.0 & $\mathcal{N}(0, 0.395^2)$ &  $-0.11\pm 0.26             $ & $-0.28\pm 0.27             $\\
         $\sigma_{\ln\lambda} $ & 0.2 & $\mathcal{U}(0.1, 0.4)$ & $0.181\pm 0.039            $ &$0.201\pm 0.039              $\\
         \hline
         \multicolumn{3}{c}{The $\eta-M_{500}-z$ scaling relation (Eq.~\ref{eq:etamass})}\\
         \hline
         %$A_{\eta}$ & 0.182 &$\mathcal{N}(0.182, 0.0073^2)$ & $0.1808\pm 0.0070          $ & $0.1839^{+0.0073}_{-0.0063}$\\
         %$B_{\eta}$ & 1.86 & $\mathcal{N}(1.86, 0.101^2)$ & $1.889\pm 0.071             $ & $1.835^{+0.061}_{-0.072} $\\
         %$\gamma_{\eta}$ & -0.83 & $\mathcal{N}(-0.83, 0.395^2)$ & $-0.68\pm 0.18$ & $-0.54^{+0.33}_{-0.29}        $\\
         $A_{\eta}$ & 0.19 &$\mathcal{N}(0.19, 0.0076^2)$ & $0.1949^{+0.0073}_{-0.0062}$&$0.1919^{+0.0077}_{-0.0066}$\\
         $B_{\eta}$ & 1.9 & $\mathcal{N}(1.9, 0.101^2)$ & $1.859^{+0.066}_{-0.076}$&$1.874^{+0.062}_{-0.073}   $\\
         $\gamma_{\eta}$ & -1.2 & $\mathcal{N}(-1.2, 0.395^2)$ & $-0.98\pm 0.26$&$-0.77^{+0.48}_{-0.41}$\\
         $\sigma_{\ln\eta}$ & 0.332 & $\mathcal{N}(0.332, 0.09^2)$ & $0.295^{+0.067}_{-0.056}   $ & $0.298^{+0.062}_{-0.055}   $\\
         \hline
         \multicolumn{3}{c}{Completeness function $C(\eta, z)$ (Eq.~\ref{eq:complfunc})}\\
         \hline
         %$\delta\log\eta_{50}$ & 0.0 & $\mathcal{U}(-0.5, 0.5)$ & $-0.0009^{+0.0064}_{-0.0072}          $ & $-0.0041\pm 0.0075          $\\
         %$s_{\log\eta-\mathrm{ex}}$ & 0.1821 & $\mathcal{U}(0.05, 0.2)$ & $0.0839\pm 0.0050          $ & $0.0834\pm 0.0052$\\
         $\delta\ln\hat\eta_{50}$ & 0.0 & $\mathcal{U}(-0.5, 0.5)$ & $-0.004\pm 0.017$ & $-0.009\pm 0.017$\\
         $s_{\ln\hat\eta}$ & 0.1819 & $\mathcal{U}(0.01, 0.46)$ & $0.193\pm 0.012$ & $0.192\pm 0.012$\\
         \hline
         \multicolumn{3}{c}{Cosmology parameters}\\
         \hline
         $\Omega_\mathrm{m}$ & 0.28 &$\mathcal{U}(0.15, 0.4)$ & $0.287\pm 0.028   $ & $0.272^{+0.024}_{-0.027}   $\\
         $h$ & 0.7 &$\mathcal{N}(0.7, 0.04^2)$ & $0.693^{+0.030}_{-0.036}            $ & $0.707^{+0.028}_{-0.031}$\\
         $\log 10^{10}\mathrm{A}_\mathrm{s}$ & 3.001 &$\mathcal{U}(1, 4)$ & $3.00\pm 0.25    $ & $2.99\pm 0.26      $\\
         $\sigma_8$ & 0.78 & - & $0.776\pm 0.031            $ &  $0.784\pm 0.033 $\\ 
         $w$ & -1.0 &$\mathcal{U}(-2, -0.5)$ & - & $-1.12\pm 0.15$\\
         \hline
    \end{tabular}
    }
\end{table}

%\subsubsection{Observable-mass scaling relations}
\subsection{Creating a RASS-MCMF mock catalog}
\label{sec:mocks}

We create mock RASS-MCMF cluster catalogues for use in forecasting the parameter constraints, adopting the \citet{Tinker08} HMF for the given cosmology and imposing a mass range $10^{12.1}$~M$_\odot <M_{500}<10^{15.5}$~M$_\odot$ and a redshift range $0.01 < z < 1.1$.  In addition, we adopt the observable--mass scaling relations presented in Section~\ref{sec:observablemass}.
The input values for the scaling relation parameters that we employ in creating the mock cluster catalogues are list in Table~\ref{tab:mocksNcosmology}, where the first column contains the parameter name and the second column the mock input value adopted.

The $\lambda$-mass relation parameters are taken to be representative of an ongoing SPT$\times$ and MARD-Y3$\times$DES cluster weak lensing analysis (Singh et al., in prep).  Because the RASS-MCMF richnesses are on average 1.5$\times$ higher than the MARD-Y3 richnesses (due to using a larger portion of the cluster luminosity function), we adjust the normalization parameter $A_\lambda$ appropriately.

For the $\eta$-mass relation parameters, we adopt best fit parameters from \cite{Chiu23} with some changes to reflect the fact that we are working with RASS count rates rather.  In particular, we absorb the parameters in the so-called bias function $b$ into the parameters of equation~(\ref{eq:etamass}).  Moreover, we adopt a characteristic scale factor between eROSITA and ROSAT count rates for the clusters of 0.117 and use that to adjust the normalization parameter $A_\eta$.

The mock sample is an X-ray existence likelihood selected sample just like the RASS-MCMF sample.  Therefore, after transforming the mass and redshift into the observed count rate $\hat\eta$ (see discussion in Section~\ref{sec:observablemass}), we use the $\hat\eta$-EXI\_ML relation
presented in Section~\ref{sec:completeness} to estimate EXI\_ML.  Doing so requires that we have an exposure time and background value for each cluster.  For these we sample the RASS-MCMF portion of the RASS sky by randomly selecting an equal area healpix pixel and using the RASS reported exposure time and background from that pixel.  We then impose the EXI\_ML threshold value of 6.5 by rejecting any cluster that falls below that limit.

The RASS-MCMF sample is also cleaned of likely random superpositions using an \fcont\ threshold that corresponds to a minimum value in $\hat\lambda(z)$.  Using the derived values $\hat\lambda(z)$ for the RASS-MCMF sample (90\% purity) or its subsamples (95\% and 99\% purity), we then reject any cluster that does not meet this observed richness threshold.  We do not add contaminating sources to the mocks.

The process of generating a mock catalog can be summarized as follows.  The first step is to construct the HOF from the HMF and observable mass scaling relations and to then integrate over the relevant ranges in observable space to estimate the expected total number of clusters.  We then draw a Poisson deviate with this expectation value, and that sets the actual number of clusters in our mock RASS-MCMF sample.
This step includes the effects of 1) the distribution of RASS exposure time and background over the RASS-MCMF sky, 2) the impact of the existence likelihood selection and 3) the impact of the \fcont\ optical cleaning.

Thereafter we cycle through the following set of steps creating the members of the mock RASS-MCMF sample:
1) We use the HMF within the specified mass and redshift ranges to randomly draw a cluster with halo mass $M_{500}$ and redshift $z$.
2) Using the $\eta-M_{500}-z$ scaling relation  (equation~\ref{eq:etamass}), we derive the count rate $\eta$ and then randomly select a RASS sky cell that has an associated exposure time and background.  With that information we predict the observed count rate $\hat\eta$.
3) Using the background, exposure time, and observed count rate we predict the existence likelihood (equation~\ref{eq:etaeximl} and impose the RASS-MCMF existence likelihood threshold EXI\_ML=6.5). 
4) We use the $\lambda-M_{500}-z$ relation to assign a richness and the sampling noise to predict an observed richness $\hat\lambda$. Then we impose the appropriate richness cut using the function $\hat\lambda_{\rm min}(z)$ that corresponds to the \fcont\ selection for the RASS-MCMF sample we are modeling.

The cosmological parameters used to create the mocks are also listed in Table~\ref{tab:mocksNcosmology}.  With this approach the total number of clusters $N_\mathrm{tot}$ from our RASS-MCMF-like 99\% purity sample ranges from 4,800 to 4,950, reflecting the Poisson variation on the expected total number of clusters for the survey.  This is the sample we employ for the forecasts described below.

\subsection{Forecast of Parameter Constraints}

We explore two different cosmogonies in our analysis: a flat $\Lambda$CDM model and a flat $w$CDM, where the dark energy equation of state parameter $w$ is a free parameter. The cosmological parameters we include are the mean dark matter density $\Omega_m$, the present epoch value of the Hubble parameter $H_0$, modeled as the dimensionless Hubble parameter $h$ where $H_0=100\,h$~km~s$^{-1}$~Mpc$^{-1}$,
the linear power spectrum amplitude $\ln(10^{10}\mathrm{A}_\mathrm{s})$
 and also the dark energy equation of state parameter $w$.  The present epoch amplitude $\sigma_8$ of matter density fluctuations on a scale of 8~$h^{-1}$~Mpc is a derived parameter.

The priors and posteriors for both models are shown in Table~\ref{tab:mocksNcosmology}.  We adopt either flat priors within specified parameter limits $\mathcal{U}(\Theta_\mathrm{min},\Theta_\mathrm{max})$ or Gaussian priors $\mathcal{N}(\left<\Theta\right>, {\rm Var}(\Theta))$ defined by their mean and variance.
For the cosmological parameters, the only Gaussian prior is for $h$ with a mean of 0.7 and a variance 0.04$^2$, which comfortably spans the recently published values \citep[e.g.,][]{Riess2019,Planck2020cosmo}. 

\begin{figure}
    \centering
    \includegraphics[width=\linewidth]{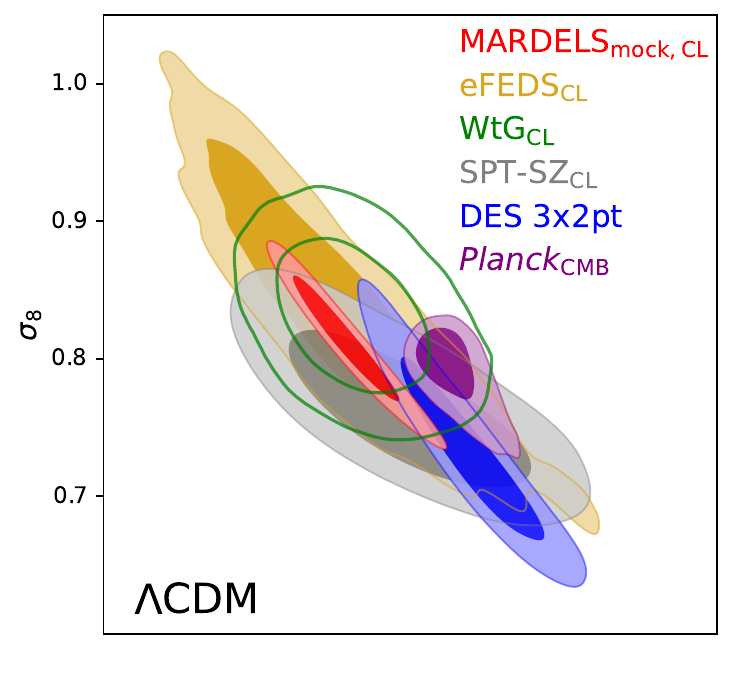}
    \vbox{\vspace{-0.85cm}
    \includegraphics[width=\linewidth]{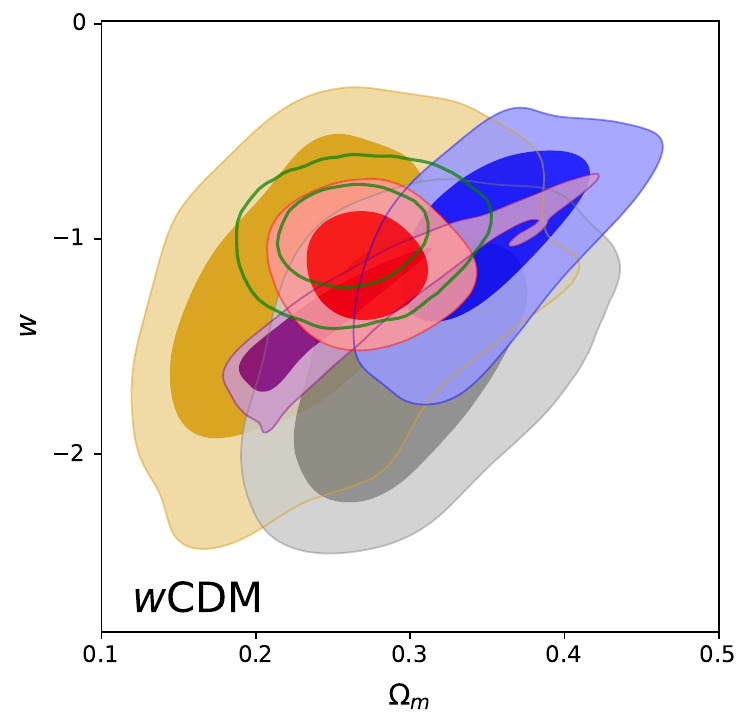}
    \vfil}
    \vspace{-0.75cm}

    \caption{Forecast cosmological parameter constraints for both a $\Lambda$CDM (\textit{top}) and a $w$CDM (\textit{bottom}) cosmogony. In both panels the contours obtained using the RASS-MCMF mock cluster catalogue (red) are compared with those obtained from different datasets: eFEDS cluster cosmology \citep[green;][]{Chiu23}, SPT-SZ cluster cosmology \citep[gray;][]{Bocquet19}, DES 3$\times$2-point constraints \citep[blue;][]{DES3x2pt} and \textit{Planck} primary CMB temperature and polarization anisotropy \citep[purple;][]{Planck2020cosmo}. The contours indicate the 1$\sigma$ and 2$\sigma$ confidence intervals.}
    \label{fig:cosmology_comparison}
\end{figure}

The priors adopted for the observable mass relation parameters reflect posteriors derived separately from an independent analysis of the MARD-Y3 and SPT cluster samples in combination with DES weak lensing (Singh et al., in prep.).
The posteriors of the observable mass and cosmological parameters for both $\Lambda$CDM and $w$CDM are consistent with the input values of the mocks at a level of $\leq 1\sigma$.  Fig.~\ref{fig:cosmology_comparison} shows the $1\sigma$ and $2\sigma$ contours of $\Omega_\mathrm{m}$ vs $\sigma_8$ and $\Omega_\mathrm{m}$ vs $w$ for our RASS-MCMF mock cluster catalogue in red, with the top panel showing the results for a flat $\Lambda$CDM cosmogony and the bottom panel showing the results for a flat $w$CDM cosmogony.

The forecast constraining power for a RASS-MCMF-like survey is in red, while a selection of previously published results including the eFEDS cluster survey \citep[yellow;][]{Chiu23}, the SPT-SZ cluster survey \citep[grey;][]{Bocquet19}, the \textit{Weighing the Giants} cluster survey \citep[WtG with green lines;][]{WtG}, the
\textit{Planck} primary CMB anisotropy using temperature and polarization \citep[\texttt{TTTEE+lowE} in purple;][]{Planck2020cosmo} and the DES 3$\times$2-point analysis \citep[blue;][]{DES3x2pt} are shown for comparison. Our results show tighter constraints than state-of-the-art cosmological analyses such as the 3$\times$2-point weak lensing analysis of DES 
%which combines information takes advantage of their relatively large solid angle ($\sim$5000~deg$^2$) and deep photometric data to estimate and combine the information 
%from three two-point correlation analyses (galaxy-galaxy, galaxy-shear and shear-shear) 
and other cluster analyses such as  those from SPT-SZ and WtG. 

It is worth noting that in the case of $w$CDM, the WtG analysis shows very similar constraints to those we forecast for the RASS-MCMF sample. The WtG contours include constraints from the assumption of constant ICM mass fraction with redshift that are impacted by tight priors adopted by the WtG team on the intrinsic evolution of the galaxy cluster ICM fraction. 
%and therefore overstate to some extent the true constraining power of the WtG sample.}  
%\textbf{It is worth noting that, in the case of $w$CDM, WtG shows very similar constraints to those we predict for the RASS-MCMF sample. The WtG contours include tight priors based on the gas mass fraction at $r\sim r_{2500}$, which in combination with external priors on mean baryon density ($\Omega_b$) cause tight constraints on $\Omega_\mathrm{m}$ and $w$.}

In the case of \textit{Planck} primary CMB constraints (bottom panel), we show the posteriors from an analysis that includes marginalization over the sum of the neutrino masses. Interestingly, for $\Lambda$CDM our results are weaker but quite competitive with \textit{Planck}, whereas the forecast RASS-MCMF posteriors are tighter than \textit{Planck} in $w$CDM.
%and $\Omega_\mathrm{m}$.

\section{Conclusions}
\label{sec:Conclusions}

In this analysis we present
%-- almost exactly 33 years after launch-- 
the RASS-MCMF cluster catalog, which has been created through a systematic search for galaxy clusters in the ROSAT 2RXS catalog \citep{Boller16} over the 25,000~deg$^2$ of the extragalactic sky covered by the Legacy Survey DR10 (Dey et al. in prep). With 8,465 clusters of galaxies, the RASS-MCMF sample is the largest ICM-selected cluster catalog to date.

Our analysis of the richness and X-ray count rates of all 2RXS counterparts indicates that the non-cluster population composes $87\pm2\%$ of the original 2RXS source list. Therefore, there are $\sim11,000$ X-ray selected galaxy clusters in the extragalactic region of this analysis. The RASS-MCMF catalog contains $\sim80$\% of the total detected cluster population in 2RXS.   Additional clusters could be extracted from the 2RXS catalog using the same method, but at the cost of increasing the contamination level of the final cluster catalog.  

The RASS-MCMF sample of 8,465 clusters presented here has an estimated 10\% contamination by residual non-cluster sources.  We also present two subsets of the RASS-MCMF sample that contain 6,924 and 5,516 clusters with 5\% and 1\% contamination, respectively (see Table~\ref{tab:catalogprop}). The full extragalactic sky coverage of the RASS-MCMF cluster sample makes it particularly interesting for further analyses.

We include spectroscopic redshifts for $\sim53\%$ of the RASS-MCMF sample using public spectroscopic galaxy redshifts. For regions with deeper spectroscopic data, such as SDSS BOSS, we find spectroscopic redshifts for $\sim93\%$ of our sample. This RASS-MCMF subsample allows cluster studies that require spectroscopic redshifts over large contiguous areas.
%which may increase in near future with new spectroscopic surveys like DESI. 
An analysis of the MCMF derived cluster photometric redshifts indicates a characteristic uncertainty of $\sigma_{\Delta z/1+z}=0.0048\pm0.0001$.

The redshift distribution of the RASS-MCMF sample peaks at $z\sim0.1$, and the new sample outnumbers any other ICM-selected cluster catalogs per redshift interval ($dN/dz$) out to $z\sim0.4$, where the SZE selected ACT-DR5 clusters start to outnumber RASS-MCMF clusters. The RASS-MCMF sample probes the galaxy group mass regime ($M_X<10^{14} M_\odot$) out to $z\sim0.15$ and therefore densely samples the cosmic web at low redshifts.

When cross-matching RASS-MCMF with three large ICM selected (X-ray or SZE selected) cluster catalogs \citep[Planck PSZ2, MCXC and ACT-DR5;][]{PSZ2,MCXC,ACTDR5}, we find between 800 and 1,000 matching clusters in each of the surveys and $\sim2,000$ matched clusters in total. 
When cross-matching the 2,815 CODEX clusters flagged as clean with the full 2RXS catalog, we find just over 70\% matches. Out of those matched sources we find $\sim85$\% also in the final RASS-MCMF sample. 
We further match RASS-MCMF with the 2RXS-based MARD-Y3 catalog \citep{Klein19} by directly matching to the same underlying 2RXS sources. We find generally good agreement between both catalogs. Sources appearing in only one of the catalogs can be explained by considering differences in catalog selection and scatter in observed richnesses.
In the matched clusters from all these catalogs, the redshifts show generally good agreement with the small number of outliers being easily explained as a simple mistake in the older catalog or a case where there are multiple optical systems along the line of sight toward the source, and the highest significance peak chosen for the RASS-MCMF cluster is not the peak selected in the other catalogs (see discussion in Section~\ref{sec:Validation}).

We use X-ray selected clusters from the eROSITA Final Equatorial-Depth Survey \citep{Brunner21,Klein22} to test and successfully confirm the purity of the RASS-MCMF sample to be 90\%.
The eROSITA X-ray survey mission \citep{Predehl21} is ongoing and has imaged the sky to greater depths than RASS.  Early expectations were that as many as 10$^5$ X-ray selected clusters could be extracted from the eROSITA dataset \citep{Merloni12}, and initial work in cluster catalog creation \citep{Brunner21,Klein22,Liu22a} and also cluster cosmology using data from an early eFEDS test survey region \citep{Chiu23} have been very encouraging.  We therefore look forward with excitement to the release of both the Russian and German parts of that rich X-ray dataset!

Our presentation of RASS-MCMF includes a cosmological forecast based on a RASS-MCMF-like mock catalog that includes various key aspects of the survey, such as X-ray existence likelihood selection, rejection of sources with low significance optical counterparts, exposure time and background variations across the sky and a realistic footprint. In addition to all these effects, we adopt realistic priors on the observable-mass scaling relation from an ongoing DES weak-lensing analysis of MARD-Y3 clusters.  The cosmological parameter uncertainties from this forecast are 0.026, 0.033 and 0.15 ($1\sigma$) on the parameters $\Omega_\mathrm{m}$, $\sigma_8$ and $w$, respectively, making RASS-MCMF$\times$DES a competitive dataset for cosmological analysis.  

\section*{Acknowledgements}
We acknowledge financial support from the MPG Faculty Fellowship program, the ORIGINS cluster funded by the Deutsche Forschungsgemeinschaft (DFG, German Research Foundation) under Germany's Excellence Strategy - EXC-2094 - 390783311, and the Ludwig-Maximilians-Universit\"at (LMU-Munich).
We gratefully acknowledge the enabling work of the DLR supported ROSAT mission and the NSF supported Legacy Surveys program.  
Detailed acknowledgements for those projects can be found in the referenced papers and are not repeated here (for LS see also https://www.legacysurvey.org/acknowledgment).
This research has made use of ``Aladin sky atlas'' developed at CDS, Strasbourg Observatory, France \citep{Aladin} and \textsc{topcat} \citep{Topcat}.

%%%%%%%%%%%%%%%%%%%%%%%%%%%%%%%%%%%%%%%%%%%%%%%%%%
\section*{Data Availability}
The catalog will be made available as online supplement as well as on CDS. Early access and additional material, such as footprint maps, additional columns or the list of point source-like rejected sources can be provided upon reasonable request.

\begin{table}
\centering
\caption{RASS-MCMF catalog column descriptions. 
}
\label{tab:catalogentries}
\begin{tabular}{|l|l|}
\hline 
 Column name & Description \\
\\
\hline
NAME	&	Cluster name	\\
RA\_OPT	&	RA of optical centre for best counterpart in degrees\\
DEC\_OPT	&	DEC of optical centre for best counterpart in degrees	\\
CENT\_TYPE	&	Type of optical centre: 1: GRZ galaxy density, \\
& 2:GRZ BCG, 3:GI galaxy density, 4:GI BCG  	\\
Z\_1	&	Photo-z best counterpart	\\
Z\_2	&	Photo-z 2nd best counterpart	\\
Z\_3	&	Photo-z 3rd best counterpart	\\
Z\_SPEC\_1	&	spectroscopic redshift best counterpart	\\
LAMBDA\_1	&	Richness best counterpart	\\
LAMBDA\_2	&	Richness 2nd best counterpart	\\
LAMBDA\_3	&	Richness 3rd best counterpart	\\
F\_CONT\_1	&	$f_\mathrm{cont}$ best counterpart	\\
F\_CONT\_2	&	$f_\mathrm{cont}$ 2nd best counterpart	\\
F\_CONT\_3	&	$f_\mathrm{cont}$ 3rd best counterpart	\\
M500\_1	&	$M_X$ for best counterpart in $\mathrm{M}_\odot$	\\
%M500\_2	&	$M_X$ for 2nd best counterpart	\\
%M500\_3	&	$M_X$ for 3rd best counterpart	\\
P\_ANY	&	$p_\mathrm{any}$ from \citet{Salvato18}	\\
P\_I	&	$P_\mathrm{I}$ from \citet{Salvato18}	\\
%DEPTH\_I	&	LS DR10 i-band depth	\\
%DEPTH\_Z	&	LS DR10 z-band depth	\\
GRZ	&	LS DR10 GRZ footprint	\\
GRZ\_N	&	LS DR10 GRZ northern footprint	\\
GI	&	LS DR10 GI footprint	\\
PSTELLAR	&	$p_\mathrm{stellar}$ from \citet{freund22}	\\
LIKELY\_STELLAR	&	Likely stellar contaminant \\
LIKELY\_QSO	&	Likely QSO contaminant \\
LG\_LAM\_MASS	&	$\log10(10^{14} \lambda/M_X)$	\\
MASKFRAC\_120	&	DR10 Mask fraction within 120 arcsec radius	\\
2RXS\_NAME	&	2RXS source name	\\
EXI\_ML	&	2RXS existance likelihood	\\
CTS	&	2RXS source counts	\\
CERR	&	2RXS uncertainty on counts	\\
RATE	&	count rate, including multiple detections in counts per second \\
2RXS\_RATE	&	2RXS count rate	in counts per second \\
2RXS\_ERATE	&	2RXS uncertainty on count rate\\
EXPOSURE	&	2RXS exposure time in seconds\\
BGR	&	2RXS background	in counts per pixel\\
2RXS\_RA\_DEG	&	2RXS RA	\\
2RXS\_DEC\_DEG	&	2RXS DEC	\\
LII	&	2RXS LII	\\
BII	&	2RXS BII	\\
EXT	&	2RXS source extend in image pixels	\\
EXTERR	&	2RXS uncertainty on source extend	\\
EXT\_ML	&	2RXS likelihood of sources being extended	\\
S\_FLAG	&	2RXS screening flag	\\
\hline
\end{tabular} 
\end{table}


\bibliographystyle{mnras}
\bibliography{MARDELS/optid_refs} % if your bibtex file is called example.bib

%%%%%%%%%%%%%%%%%%%%%%%%%%%%%%%%%%%%%%%%%%%%%%%%%%
\appendix

\section{Planck-PSZ2 clusters with potentially incorrect redshifts}
In the table below we list RASS-MCMF matches to Planck PSZ2 clusters where the redshift listed in the PSZ2 catalog is likely incorrect.

\begin{table*}
    \caption{PSZ2 clusters with possible incorrect redshifts. We list cluster name (PSZ2 name), position and redshifts from PSZ2 (RA$_\mathrm{PSZ2}$,DEC$_\mathrm{PSZ2}$,z$_\mathrm{PSZ2}$) as well as 2RXS X-ray position (RA$_\mathrm{2RXS}$,DEC$_\mathrm{2RXS}$) and MCMF based measurements of redshift ($z$), richness ($\lambda$) and \fcont\ for the best and second best counterpart. We finally provide a comment to each cluster.}
    \label{app:PSZphotozoff}
    \centering
        \resizebox{\linewidth}{!}{%
\begin{tabular}{|l|r|r|r|l|r|l|r|r|r|r|r|r|}
\hline
  \multicolumn{1}{|c|}{PSZ2 NAME} &
  \multicolumn{1}{c|}{RA$_\mathrm{PSZ2}$} &
  \multicolumn{1}{c|}{DEC$_\mathrm{PSZ2}$} &
 % \multicolumn{1}{c|}{Other name} &
  \multicolumn{1}{c|}{z$_\mathrm{PSZ2}$} &
%  \multicolumn{1}{c|}{COMMENT_PSZ} &
  \multicolumn{1}{c|}{RA$_\mathrm{2RXS}$} &
  \multicolumn{1}{c|}{DEC$_\mathrm{2RXS}$} &
  \multicolumn{1}{c|}{$z_1$} &
  \multicolumn{1}{c|}{$z_2$} &
  \multicolumn{1}{c|}{$\lambda_1$} &
  \multicolumn{1}{c|}{$\lambda_2$} &
  \multicolumn{1}{c|}{$f_\mathrm{cont,1}$} &
  \multicolumn{1}{c|}{$f_\mathrm{cont,2}$} &
 % \multicolumn{1}{c|}{$z_\mathrm{spec}$} &
 % \multicolumn{1}{c|}{Offset [arcsec]} &
  \multicolumn{1}{c|}{Comment} \\
\hline
 PSZ2 G091.40-51.01 & 353.5028 & 7.0573 &  0.099 &   353.4740 & 7.0703 & 0.295 & 0.551 & 160.2 & 14.6 & 0.00 & 0.94  & RASS-MCMF correct ($z_\mathrm{spec}=0.2955$) \\ 
 PSZ2 G109.86+27.94 & 275.8330 & 78.3893 &    0.4 &   275.7792 & 78.3684 & 0.669 & 0.045 & 233.3 & 6.9 & 0.00 & 0.31 &    RASS-MCMF likely  correct ($\lambda=233$ system) \\ 
 PSZ2 G181.71-68.65 & 31.6490 & -14.8800 &   0.1529 &  31.6192 & -14.8970 & 0.293 & 0.240 & 181.7 & 40.8 & 0.00 & 0.10 &    RASS-MCMF correct, PSZ redshfit from foregr. spiral \\ 
%% PSZ2 G254.96+55.88 & 168.2173 & 2.4837 & RXC J1113.3+0231 & 0.078 & SZ projection: z=0.08 \& z=0.27 & 168.2122 & 2.4988 & 0.265 & 0.069 & 122.3 & 37.8 & 0.00 & 0.02 & 0.2655 & 57 & RASS-MCMF correct for 2RXS source, PSZ detection points to nearby ACO 1205  \\ 
 PSZ2 G281.09-42.51 & 56.2994 & -66.5015 &    0.38 &   56.3505 & -66.5000 & 0.557 & 0.130 & 153.1 & 6.4 & 0.00 & 0.59 &    RASS-MCMF likely correct ($\lambda=153$ system) \\ 
 PSZ2 G287.00-35.24 & 68.1017 & -74.1685 &   0.43 &   68.0470 & -74.1689 & 0.166 & 0.501 & 43.3 & 63.1 & 0.05 & 0.14 &    two clusters along the line of sight, PSZ redshift likely merges both clusters \\ 
%% PSZ2 G341.19-36.12 & 308.0656 & -56.4352 & SPT-CLJ2032-5627  & 0.284 & SZ projection: z=0.14 \& z=0.28 & 307.9994 & -56.4391 & 0.129 & 0.282 & 74.3 & 95.1 & 0.00 & 0.01 & 0.0000 & 133 & Two clusters in RASS-MCMF, PSZ detection corresponds to the other RASS-MCMF source \\ 
%% PSZ2 G351.76-54.71 & 334.6925 & -45.2857 & SPT-CLJ2218-4519  & 0.61 & Point sources at: 545GHz & 334.7528 & -45.2295 & 0.184 & 0.250 & 26.5 & 4.1 & 0.16 & 0.82 & 0.0000 & 254 & Likely chance match, PSZ detection corresponds to other cluster \\ 
\hline
\end{tabular}
}
\end{table*}

\section{Sources from the NEP survey likely misclassified as clusters}
In table~\ref{app:NEPmisc} we list 18 sources from the NEP survey \citep{Henry_2006} classified as clusters that might not be real clusters after visual inspection. The first ten sources further have a counter part in the Million Quasars Catalog \citep{Flesch21}.

\begin{table*}
    \caption{List of 2RXS matches to the ROSAT NEP survey that fail visual inspection. 2RXS sources that do have a match to the Million Quasars Catalog \citep{Flesch21} are listed first with names of the corresponding QSO, type and redshift listed, while sources without match to the Million Quasars Catalog are appended. We provide source name and position from the 2RXS catalog (2RXS Name, 2RXS\_RA, 2RXS\_DEC), redshift ($z$), richness ($\lambda$) and \fcont\ from MCMF measurements. Similar we provide source position and names of the matched NEP source}
    \label{app:NEPmisc}
    \centering
        \resizebox{\linewidth}{!}{%
\begin{tabular}{|l|r|r|r|r|r|r|r|r|l|r|l|l|r|}
\hline
  \multicolumn{1}{|c|}{2RXS NAME} &
  \multicolumn{1}{c|}{2RXS\_RA} &
  \multicolumn{1}{c|}{2RXS\_DEC} &
  \multicolumn{1}{c|}{$z_1$} &
  \multicolumn{1}{c|}{$\lambda_1$} &
  \multicolumn{1}{c|}{\fcont} &
  %\multicolumn{1}{c|}{P\_ANY} &
  \multicolumn{1}{c|}{NEP\_RA} &
  \multicolumn{1}{c|}{NEP\_DEC} &
  \multicolumn{1}{c|}{NEP\_NAME} &
  \multicolumn{1}{c|}{z\_NEP} &
  \multicolumn{1}{c|}{QSO\_Name} &
  \multicolumn{1}{c|}{QSO\_Type} &
  \multicolumn{1}{c|}{QSO\_z} \\
\hline
2RXS J180606.6+681308 & 271.5276 & 68.2191 & 0.269 & 9.2 & 0.65 &  271.5275 & 68.2189 & RX J1806.1+6813 & 0.303 & WISEA J180609.00+681309.6 & qX & 0.3\\
 2RXS J171640.0+641048 & 259.1669 & 64.1801 & 0.249 & 18.7 & 0.35 &  259.1654 & 64.1764 & RX J1716.6+6410 & 0.251 & WISEA J171636.31+641112.3 & q & 0.2\\
 2RXS J180844.0+655705 & 272.1835 & 65.9516 & 0.495 & 5.5 & 1.00 &  272.1817 & 65.9514 & RX J1808.7+6557 & 0.246 & J180843.17+655705.4 & X & 0.3\\ 
 2RXS J172839.2+704105 & 262.1634 & 70.6848 & 0.276 & 2.4 & 0.90 &  262.1646 & 70.6847 & RX J1728.6+7041 & 0.551 & RXS J17286+7041 & QRX & 0.551\\
 2RXS J180732.2+642919 & 271.8846 & 64.4886 & 0.220 & 1.9 & 0.90 &  271.8846 & 64.4881 & RX J1807.5+6429 & 0.239 & 3HSPJ180732.2+642926 & BRX & 0.239\\ 
 2RXS J182237.8+664132 & 275.6575 & 66.6924 & 0.726 & 68.0 & 0.49 &  275.6558 & 66.6914 & RX J1822.6+6641 & 0.089 & J182237.52+664126.0 & X & \\ 
 2RXS J175857.7+652057 & 269.7407 & 65.3494 & 0.283 & 4.8 & 0.83 &  269.7400 & 65.3494 & RX J1758.9+6520 & 0.365 & WISEA J175856.74+652106.5 & qX & 0.5\\
 2RXS J175406.5+645201 & 268.5272 & 64.8672 & 0.249 & 7.1 & 0.69 &  268.5221 & 64.8669 & RX J1754.0+6452 & 0.246 & CGRaBS J1754+6452 & QRX & 0.977\\
 2RXS J174949.4+682319 & 267.4559 & 68.3886 & 0.522 & 9.0 & 1.00 &  267.4575 & 68.3875 & RX J1749.8+6823 & 0.051 & KUG 1750+683A & NRX & 0.051\\
 2RXS J183917.7+701820 & 279.8239 & 70.3058 & 0.103 & 6.6 & 0.55 &  279.8225 & 70.3056 & RX J1839.2+7018 & 0.230 & WISEA J183917.18+701823.7 & qX & 0.3\\
 2RXS J172124.1+673313 & 260.3508 & 67.5537 & 1.491 & 3.5 & 0.99 &  260.3525 & 67.5539 & RX J1721.4+6733 & 0.0861 & & &  \\ 
 2RXS J172411.5+700027 & 261.0481 & 70.0078 & 0.325 & 5.2 & 0.86 &  261.0483 & 70.0075 & RX J1724.1+7000 & 0.0386 & & &   \\ 
 2RXS J174516.4+655617 & 266.3187 & 65.9382 & 0.612 & 22.9 & 0.89 &  266.3175 & 65.9381 & RX J1745.2+6556 & 0.608  & & &  \\ 
 2RXS J175130.6+701415 & 267.8779 & 70.2378 & 0.446 & 18.4 & 0.69 &  267.8779 & 70.2256 & RX J1751.5+7013 & 0.4925 & & &  \\ 
 2RXS J175211.8+652222 & 268.0494 & 65.3730 & 0.072 & 7.8 & 0.40 &  268.0500 & 65.3728 & RX J1752.2+6522 & 0.3923 & & &  \\ 
 2RXS J180416.1+672922 & 271.0672 & 67.4896 & 0.040 & 1.0 & 0.88 &  271.0650 & 67.4892 & RX J1804.2+6729 & 0.0617 & & &  \\ 
 2RXS J181119.1+644738 & 272.8297 & 64.7941 & 0.179 & 5.0 & 0.72 &  272.8296 & 64.7933 & RX J1811.3+6447 & 0.451 & & &  \\ 
 2RXS J181208.3+635336 & 273.0346 & 63.8935 & 0.046 & 6.3 & 0.36 &  273.0350 & 63.8931 & RX J1812.1+6353 & 0.5408 & & &  \\ 

\hline
\end{tabular}
}
\end{table*}

% Don't change these lines
\bsp	% typesetting comment
\label{lastpage}
\end{document}